\documentclass[smallextended]{svjour3} 

\def\BibTeX{{\rm B\kern-.05em{\sc i\kern-.025em b}\kern-.08em
    T\kern-.1667em\lower.7ex\hbox{E}\kern-.125emX}}
    
\journalname{Empirical Software Engineering}

\usepackage{orcidlink} 
\usepackage{multirow}%
\usepackage{mathrsfs}%
\usepackage{manyfoot}%
\usepackage{array}
\usepackage{amsfonts}
\usepackage{cite}
\usepackage{algorithmic}
\usepackage{graphicx}
\usepackage{textcomp}
\usepackage{xcolor}
\usepackage{listings}
\usepackage{sansmath}
\usepackage{hyperref}
\usepackage{tikz}
\usepackage{tcolorbox}
\usepackage{nicefrac}
\usepackage{balance}
\usepackage{lipsum}
\usepackage{subcaption}
\usepackage{float}
\usepackage{booktabs}
\usepackage{xspace}
\usepackage{tikz}
\usepackage{adjustbox}
\usepackage{subcaption}
\usepackage{amsmath}
\usepackage{wrapfig}
\usepackage{physics}
\usepackage{tablefootnote}
\usepackage{threeparttable}
\usepackage{caption} 
\usepackage{lineno}
\usepackage[normalem]{ulem}
\usepackage{makecell}
\usepackage{xcolor}
\usepackage{colortbl}
\usepackage{graphicx}
\usepackage{amsmath,amssymb}

\usepackage[vlined, boxruled, linesnumbered] {algorithm2e}
\SetKwComment{Comment}{$\triangleright$\ }{}
\SetCommentSty{itshape}
\newboolean{showcomments}
\setboolean{showcomments}{true}
\ifthenelse{\boolean{showcomments}}
{
	\definecolor{myyellow}{RGB}{255, 228, 26}
	\definecolor{myblue}{RGB}{50, 50, 220}
	\newcommand{\nb}[2]{
		{\sf
			\fcolorbox{myyellow}{yellow}{\scriptsize\textbf{#1}}%
			$\blacktriangleright$%
			{\color{myblue}\fontsize{9pt}{9pt}\selectfont\textbf{#2}}%
		}%
	}
}
{
	\newcommand{\nb}[2]{}
}

\newcommand{\ssim}[0]{\texttt{SingleSim}\xspace}

\newcommand{\msim}[0]{\texttt{MultiSim}\xspace}

\newcommand{\davetwo}{\mbox{DAVE-2}\xspace} %
\newcommand{\vit}{\mbox{ViT}\xspace} %
\newcommand{\ticp}{\mbox{TCP}\xspace} %

\newcommand{\dss}[0]{\texttt{DSS}\xspace}

\usepackage{soul}

\newcommand{\head}[1]{\noindent\textbf{#1.}}

\newcounter{commentnumber}

\newcommand{\blue}[1]{\begingroup\color{black}#1\endgroup}

\newcommand{\changed}[1]{\begingroup\color{black}#1\endgroup}

\begin{document}

\pagenumbering{arabic} 
\pagestyle{plain}

\title{
Simulator Ensembles for Trustworthy Autonomous Driving Systems Testing
}

\author{
Lev Sorokin \textsuperscript{1} \orcidlink{0009-0003-1162-6252} \and
Matteo Biagiola \textsuperscript{2} \orcidlink{0000-0002-7825-3409} \and 
Andrea Stocco \textsuperscript{1,3} \orcidlink{0000-0001-8956-3894}
}

\institute{
    \textsuperscript{1} Technical University of Munich, Munich, Germany. Email: lev.sorokin@tum.de, andrea.stocco@tum.de \\
    \textsuperscript{2} University of St. Gallen and Università della Svizzera italiana, St. Gallen / Lugano, Switzerland. Email: matteo.biagiola@{unisg,usi}.ch \\
    \textsuperscript{3} fortiss GmbH, Munich, Germany. Email: stocco@fortiss.org \\
}

\titlerunning{Mitigating Simulator-Specific Failures}
\authorrunning{Sorokin et al.}

\date{Received: date / Accepted: date
}

\maketitle

\begin{abstract}
Scenario-based testing with driving simulators is extensively used to identify failing conditions of automated driving assistance systems (ADAS) and reduce the amount of in-field road testing.
However, existing studies have shown that repeated test execution in the same as well as in distinct simulators can yield different outcomes, which can be attributed to sources of flakiness or different implementations of the physics, among other factors. 

In this paper, we present \texttt{MultiSim}, a novel approach to multi-simulation ADAS testing based on a search-based testing approach that leverages an ensemble of simulators to identify failure-inducing, simulator-agnostic test scenarios. During the search, each scenario is evaluated jointly on multiple simulators. Scenarios that produce consistent results across simulators are prioritized for further exploration, while those that fail on only a subset of simulators are given less priority, as they may reflect simulator-specific issues rather than generalizable failures.

\changed{Our empirical study, which involves testing three lane-keeping ADAS with increasing complexity} on different pairs of three widely used simulators, demonstrates that \texttt{MultiSim} outperforms single-simulator testing by achieving, on average, a higher rate of simulator-agnostic failures by \changed{66\%}. Compared to a state-of-the-art multi-simulator approach that combines the outcome of independent test generation campaigns obtained in different simulators, \texttt{MultiSim} identifies on average up to \changed{$3.4\times$ more simulator-agnostic failing tests and higher failure rates.}

To avoid the costly execution of test inputs on which simulators disagree, we propose an enhancement of \msim that leverages surrogate models to predict simulator disagreements and bypass test executions. Our results show that utilizing a surrogate model during the search does not only retain the average number of valid failures but also improves its efficiency in finding the first valid failure. These findings indicate that combining an ensemble of simulators during the search is a promising approach for the automated cross-replication in ADAS testing. 

\end{abstract}

\noindent \textbf{Keywords.}
Search-based software testing, scenario-based testing, autonomous driving, testing deep learning systems, simulator-agnostic

\section{Introduction} \label{sec:intro}

Advanced Driver Assistance Systems (ADAS) rely on perception systems like cameras and LiDAR, powered by deep neural networks (DNNs), for real-time tasks such as lane-keeping, object detection and image segmentation. While effective, these systems must operate reliably across diverse environments, yet training data cannot cover all possible scenarios~\cite{survey-lei-ma,lou2022testing}. Consequently, ADAS may encounter unseen inputs post-deployment, making DNNs highly sensitive to variations in road shapes, lighting, and noise. These discrepancies can lead to ADAS prediction errors, misclassifications, and inaccurate segmentations, which impact vehicle decision-making and compromise safety~\cite{Cerf:2018:CSC:3181977.3177753,2025-Lambertenghi-ICST}.

Validating the safety of ADAS through virtual testing with scenario-based simulation is the default option for companies~\cite{survey-lei-ma,Cerf:2018:CSC:3181977.3177753}. Simulators enable developers to quickly prototype ADAS and evaluate them across a wide range of challenging scenarios.
In the literature, researchers have proposed automated testing techniques to expose failing conditions and corner cases~\cite{asfault,Abdessalem-ICSE18,2020-Riccio-FSE,nejati2023reflections, Kluck19Nsga2ADAS,matinnejad2015,Zeller17SBST,marko2019}, using various open-source ADAS simulators, such as CARLA~\cite{carla}, BeamNG~\cite{beamng}, and Udacity~\cite{udacity-simulator}, or commercial close-source solutions, such as Siemens PreScan~\cite{prescan}, ESI Pro-SiVIC~\cite{pro-sivic}, and PTV VISSIM~\cite{vissim}. 

However, the result of a test execution through a simulation environment only approximates the actual test outcome in the real world. Indeed, multiple studies have shown that simulation-based testing might produce inconsistent results both within simulator, i.e., when the same test scenario is executed multiple times on the same simulator~\cite{AminiFlaky2024,birchler2023machine}, and across simulators~\cite{Borg21CrossSimTesting,matteoMaxibon2024}. This phenomenon may be due to test flakiness, and it is characterized by multiple runs of the same test exhibiting non-deterministic behavior (i.e., tests pass or fail non-deterministically). Test flakiness has been studied extensively in the software testing literature~\cite{parry2022survey}, including ADAS testing. A study by Amini et al.~\cite{AminiFlaky2024} shows that test flakiness is quite common also for simulation-based testing of ADAS, potentially leading to a distrust in virtual testing, as also reported in the field of robotic simulation~\cite{icst-survey-robotics}. Possible causes of flakiness stem from uncertainties in the simulation environment to timing and synchronization in the interaction between the ADAS and the simulator~\cite{AminiFlaky2024}. In general, such sources of non-determinism are unknown and difficult to control, making the results of automated testing techniques, i.e., the failure-inducing test cases, unreliable.

Despite the significance of the problem, very few solutions have been proposed to mitigate test flakiness in simulation-based testing (beyond discarding flaky tests~\cite{birchler2023machine}). Amini et al.~\cite{AminiFlaky2024} used machine-learning classifiers to predict the flakiness of an ADAS test using few test runs. Another way to mitigate test flakiness is by combining the outcome of multiple simulators when performing a testing campaign. In particular, Biagiola et al.~\cite{matteoMaxibon2024}, proposed a multi-simulator approach to approximate the outcome of a test case on a digital twin. Their framework named \textit{digital siblings} (\dss henceforth) executes two independent test generation algorithms on two simulators, generating two feature maps that characterize test cases executed on such simulators. Through the operations of migration and merge, the framework outputs a combined feature map that prioritizes \textit{agreements} across simulators, i.e., when the same test case has the same outcome on different simulators. The intuition is that we can expect test cases where the two simulators agree to be more reliable, and hence less flaky, than those where the two simulators disagree. Although their results show that the combination of the two simulators is able to predict the failures of the digital twin better than each individual one, the drawback of the digital siblings framework is that test case execution outcomes on the two simulators are merged as a post-processing step. This way, the test generation algorithm runs the risk of evolving flaky tests as each test is evaluated only on a single simulator during the search, making the whole process inefficient.

In this paper we present a novel approach named \msim, which incorporates multiple simulators as an ensemble directly as part of the test generation process. In particular, we cast the testing problem as a multi-objective optimization problem, where we define, for each simulator, a fitness function that evaluates the quality of the ADAS under test. 
\changed{In this way, we treat simulator disagreement as a first-class citizen during the optimization process.} 
By minimizing the fitness values obtained from executing each test case in multiple simulators and their distance, we direct the search towards failure-inducing tests that are \textit{simulator-agnostic}, i.e, reliable, tests since the respective outcomes on the two simulators are close to each other.

\changed{We evaluate our \msim approach with three lane-keeping ADAS based on convolutional DNNs\cite{nvidia-dave2}, transformers~\cite{2025-Baresi-ICSE} and end-to-end trajectory/control prediction~\cite{tcp}, on three different and widely used simulators, namely Udacity, Donkey and BeamNG~\cite{asfault,2021-Jahangirova-ICST,2020-Riccio-FSE,2020-Stocco-ICSE,zohdinasab2021deephyperion,AminiFlaky2024}.}
In particular, we compare \msim with a single simulator approach, as well as with the \dss framework in terms of effectiveness, i.e., the number of simulator-agnostic failures each approach triggers given a fixed search budget, and efficiency, i.e., how quickly each approach generates the first simulator-agnostic failure. 

Our results show that \msim identifies on average 70\% valid failures across all simulator configurations, outperforming \dss with 65\% and single simulator-based testing with 47\%. Regarding the rate of simulator-agnostic failures, BD (i.e., BeamNG combined with Donkey) outperforms both single-simulator testing and \dss-based testing, except for \dss-BD, where it has a similar validity rate but identifies a significantly higher number of failures. Using an ensemble of simulators during search-based testing demonstrates efficiency comparable to testing with only one simulator. Compared to \dss, in the BD configuration, it achieves significantly better results, while in another comparison, i.e., BU (BeamNG combined with Udacity) is outperformed by \dss-BU. Finally, integrating a Random Forest classifier to predict simulator-related disagreements into \msim improves the effectiveness, leading to a higher median of simulator-agnostic failures, as well as a lower standard deviation, compared to running the search without the classifier. Similarly, in terms of efficiency, we can identify a smaller median and variation in the search budget required for finding the first valid failure.

Our paper makes the following contributions:
    
\begin{itemize}
    \item \textbf{Approach.} A novel search-based test generation approach called \msim that integrates multiple simulators to identify simulator-agnostic failures for ADAS.
     \item \textbf{Evaluation.} An extensive empirical study using three simulators and three different approaches showing that \msim is effective at generating simulator-agnostic failures for \changed{three} state-of-the-art lane-keeping ADAS. To encourage open research, our test generation approach and experimental data are available~\cite{replication-package}.
\end{itemize}

The outline of the paper is as follows: In Section 2, we present the preliminaries. In Section 3, we introduce the problem statement and the notion of simulator-agnostic failures. In Section 4, we present our test generation approach, followed by our validation strategy. In Section 5, we evaluate our testing approach in comparison with single-simulator based testing and a default multi-simulator based testing approach. In Section 6, we discuss the insights and clarifications regarding our approach. In Section 7, we outline the most important threats regarding the validity of our results. In Section 8, we present related work.  We finally conclude in Section 9, with a summary and insights into future work.


\section{Background}\label{sec:background}

\subsection{Simulation-based ADAS Testing}
\label{sec:simulation-based-testing}

We consider Level 2 ADAS, as classified by the National Highway Traffic Safety Administration (NHTSA)~\cite{nhtsa}, which perform vision-based perception tasks using data gathered by camera sensors of a vehicle. 
\changed{Despite the adoption of Level 2 ADAS in many commercial vehicles, their reliability remains a concern, as evidenced by numerous recent crash reports~\cite{NHTSA-level2-crashes} and large scale real-world validation experiments~\cite{Opletal2025ChinaADAS}. 
Although Levels 3 and 4 ADAS have been proposed~\cite{baiduapolloscapes}, their real-world deployment remains highly constrained. Consequently, addressing the limitations of Level 2 systems is crucial for advancing to higher levels of autonomy. Particularly, in this paper we focus on ADAS that learn the \textit{lane-keeping} functionality, critical component for the safe operation of self-driving vehicles, from human-labeled driving data.}

In the early stages of development, ADAS undergo model-level testing~\cite{2020-Riccio-EMSE}. This involves evaluating performance metrics such as accuracy, mean squared error (MSE), and mean absolute error (MAE) on an unseen test set, i.e., a dataset not used during training. This form of testing is analogous to unit testing in traditional software development, helping test engineers identify inadequately trained models~\cite{2020-Humbatova-ICSE}.

Following model-level testing, ADAS are subjected to system-level testing~\cite{2021-Haq-EMSE,Codevilla,2023-Stocco-EMSE}. This phase assesses the impact on the entire decision-making process of an ADAS, ensuring that its predictions align with expected vehicle behavior. A typical system-level test, i.e., test scenario, involves generating a one-lane or two-lane road, each defined by a starting and an endpoint, with varying length, curvature, and number of turns~\cite{2025-Ali-ICSEW}.
While driving, the ADAS processes input images and generates steering commands. System failures occur when the vehicle deviates from system requirements, such as violating safety constraints, e.g., driving off the road or causing harm to other vehicles, the environment, or pedestrians~\cite{2020-Stocco-ICSE}. These failures often stem from errors in the perception component~\cite{2024-Lambertenghi-ICST,2025-Lambertenghi-ICST}. The testing objective is therefore to identify road topologies where the ADAS fails to maintain the vehicle in lane, either by driving off-road (for one-lane scenarios) or crossing the opposite lane (for two-lane scenarios).

System-level ADAS testing is primarily conducted in simulation environments using software-in-the-loop testing to ensure safety and minimize costs. This approach enables safe measurement, analysis, characterization, and reproduction of driving failures. Well-established ADAS simulation platforms include industrial platforms such as Siemens PreScan~\cite{prescan} or ESI Pro-SiVIC~\cite{pro-sivic}, whereas open-source solutions widely adopted by researchers include Udacity~\cite{udacity-simulator}, Donkey Car's sdsandbox~\cite{donkeycar} (hereafter referred to as Donkey, for simplicity), and BeamNG~\cite{beamng}.

\subsection{Search-based Software Testing}\label{sec:sbt}

Our framework uses the concepts behind search-based testing (SBST), a testing technique where the testing problem is modeled as an optimization problem to be solved with metaheuristic optimization techniques~\cite{McMinn2011SBT}. SBST is defined as follows:

\begin{definition}
A search-based testing problem $P$ is defined as a tuple $P = (T, D, F, O)$, where

\begin{itemize}
    \item $T$ is the system under test.
    
    \item $D \subseteq \mathbb{R}^n$ is the search domain, where $n$ is the \textit{dimension} of the search space. The vector $\mathbf{x} =(x_1, \ldots, x_n) \in D$ is called test input. 
    
    \item $F$ is the vector-valued fitness function defined as $F: D \mapsto \mathbb{R}^m, F(\mathbf{x}) = (f_1(\mathbf{x}),\ldots, f_m(\mathbf{x}))$, where $f_i$ is a scalar fitness function (or fitness function, for short) that assigns a quantitative value to each test input and $\mathbb{R}^m$ is the \textit{objective space}, and $m$ corresponds to the number of fitness functions. A fitness function evaluates how \textit{fit} a test input is, assigning a \emph{fitness value} to it.
    
    \item $O$ is the oracle function, $O : \mathbb{R}^m \mapsto \lbrace 0,1 \rbrace $, which evaluates, given the objective space of the fitness functions, whether a test passes or fails. A test that fails is called \emph{failure-inducing}.
\end{itemize}

\end{definition}

For instance, a search based testing problem for the ADAS system from \autoref{sec:simulation-based-testing} is defined as follows: roads represent test inputs passed to the ADAS and the execution of the ADAS on a road is performed in a simulator yielding location traces. Location traces are used by the fitness function to calculate to each executed test input the maximal distance of the vehicle to the center line. The oracle function specifies that an execution is failing when the vehicle drives off the lane when the fitness function value is above a predefined threshold.

In our approach, solving such a search-based testing problem requires the definition of multiple fitness functions. We use the concepts of multi-objective optimization to solve the problem.

\begin{definition}
In search-based testing, a multi-objective optimization (MOO) problem is defined as:

$$
\min_{\mathbf{x} \in X} F(\mathbf{x}) = (f_1(\mathbf{x}), \ldots, f_m(\mathbf{x}))
$$

\noindent
where $f_i$ is a scalar fitness function and $X \subseteq D$ is called the feasible solution set. 
In general, a set of equality and inequality constraints is defined, which have to be satisfied by solutions in $X$. A solution $\mathbf{x}$ with $F(\mathbf{x}) = (v_1,\ldots, v_m)$ is said to \textit{dominate} another solution $\mathbf{x}'$ with $F(\mathbf{x}') =(v_1',\ldots, v_m') \iff \exists v_i. (v_i < v_i') \wedge \forall v_j.(v_j \leq v_j')$, i.e., $\mathbf{x}$ is superior to $\mathbf{x}'$ in at least one fitness value and at least as good in all other fitness values.
Consequently, a solution $\mathbf{x}$ is called Pareto optimal if no solution exists that dominates it.
The set of all Pareto optimal solutions is called Pareto set $PS$.

\end{definition}


\section{Problem Definition}
\label{sec:motivation}

\begin{figure}[t] 
  \begin{subfigure}[h]{0.3\columnwidth} 
      \centering
      \includegraphics[trim={0cm 0cm 0cm 2.5cm},clip, scale=0.21]{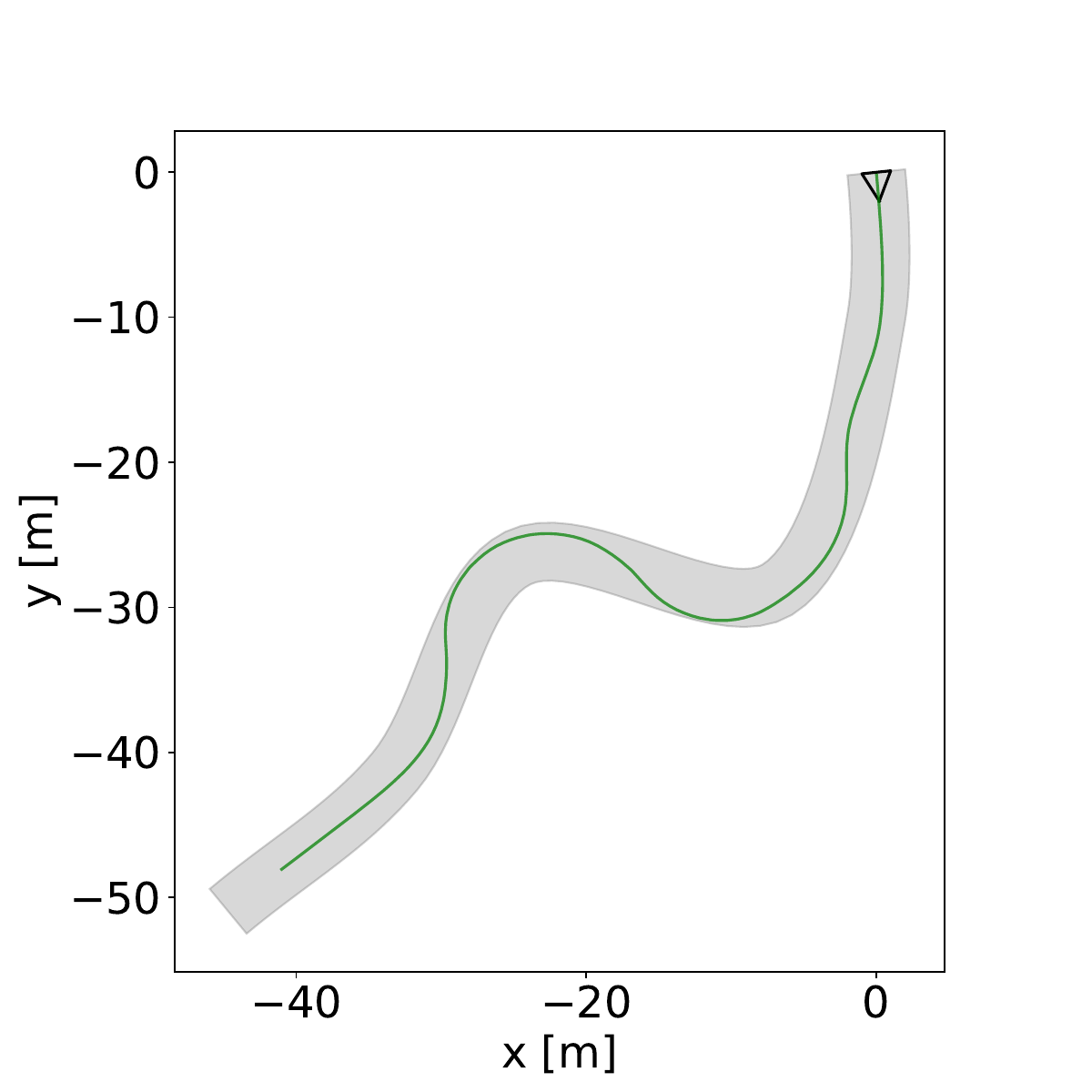}
      \caption{Udacity.}
      \label{fig:flakiness_low}
  \end{subfigure}%
  \hfill
  \begin{subfigure}[h]{0.3\columnwidth} 
      \centering
      \includegraphics[trim={0cm 0cm 0cm 2.5cm},clip,scale=0.21]{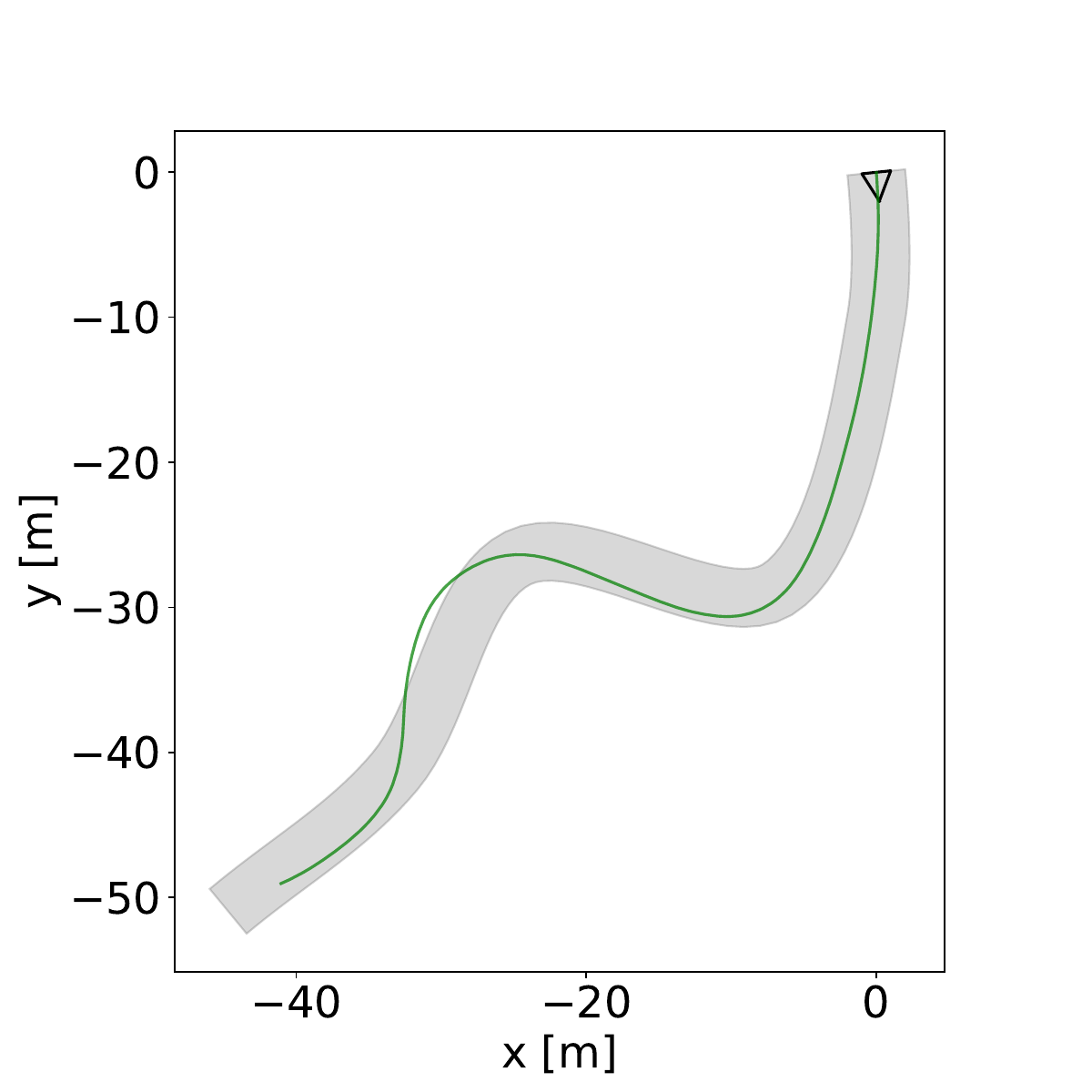}
      \caption{Donkey.}
      \label{fig:flakiness_high}
  \end{subfigure}
  \hfill
  \begin{subfigure}[h]{0.3\columnwidth} 
      \centering
      \includegraphics[trim={0cm 0cm 0cm 2.5cm},clip,scale=0.21]{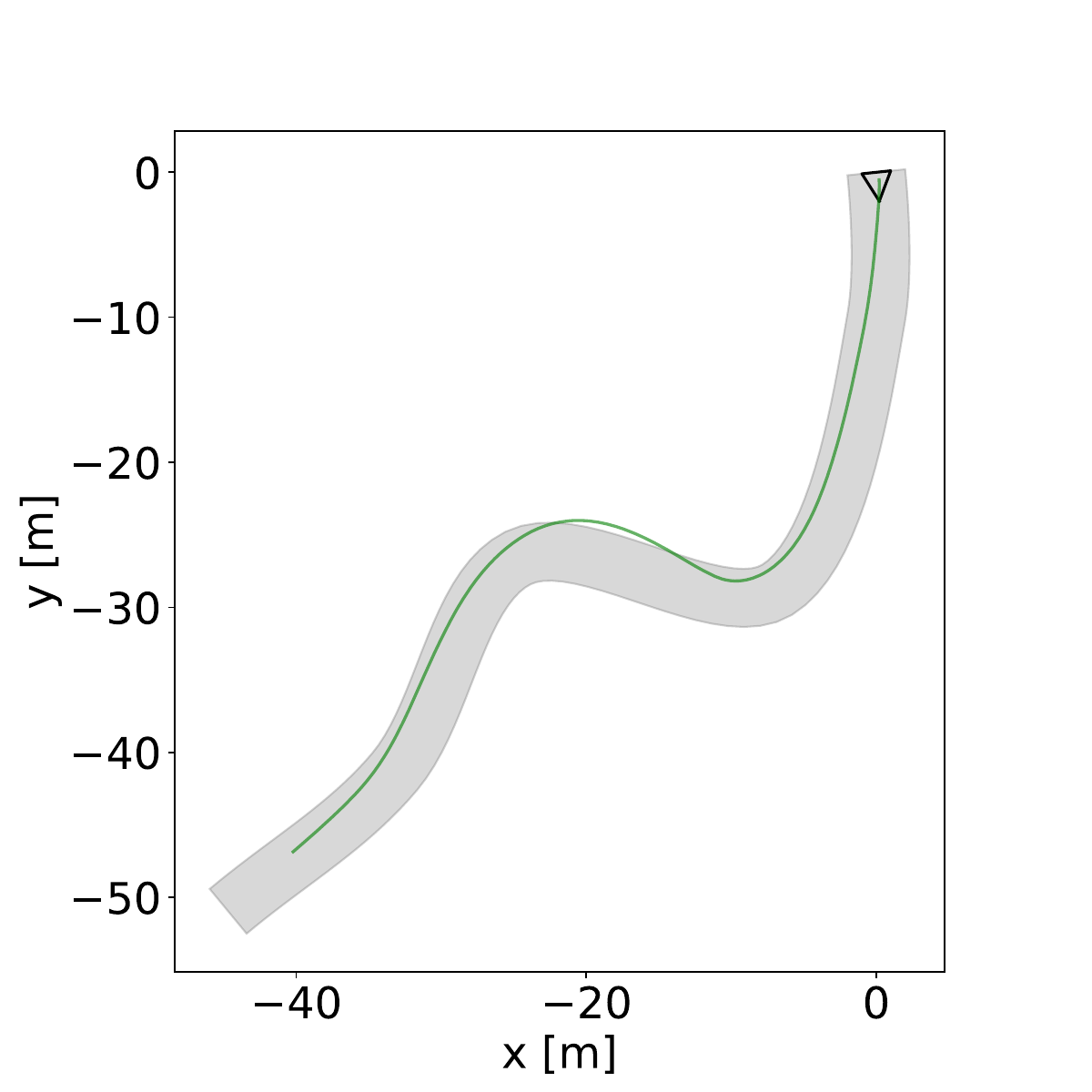}
      \caption{BeamNG.}
      \label{fig:flakiness_high_bng}
  \end{subfigure}
        \hfill
    \begin{subfigure}[h]{0.3\columnwidth}
        \centering
        \includegraphics[trim={4cm 4cm 2.5cm 4cm},clip,scale=0.107]{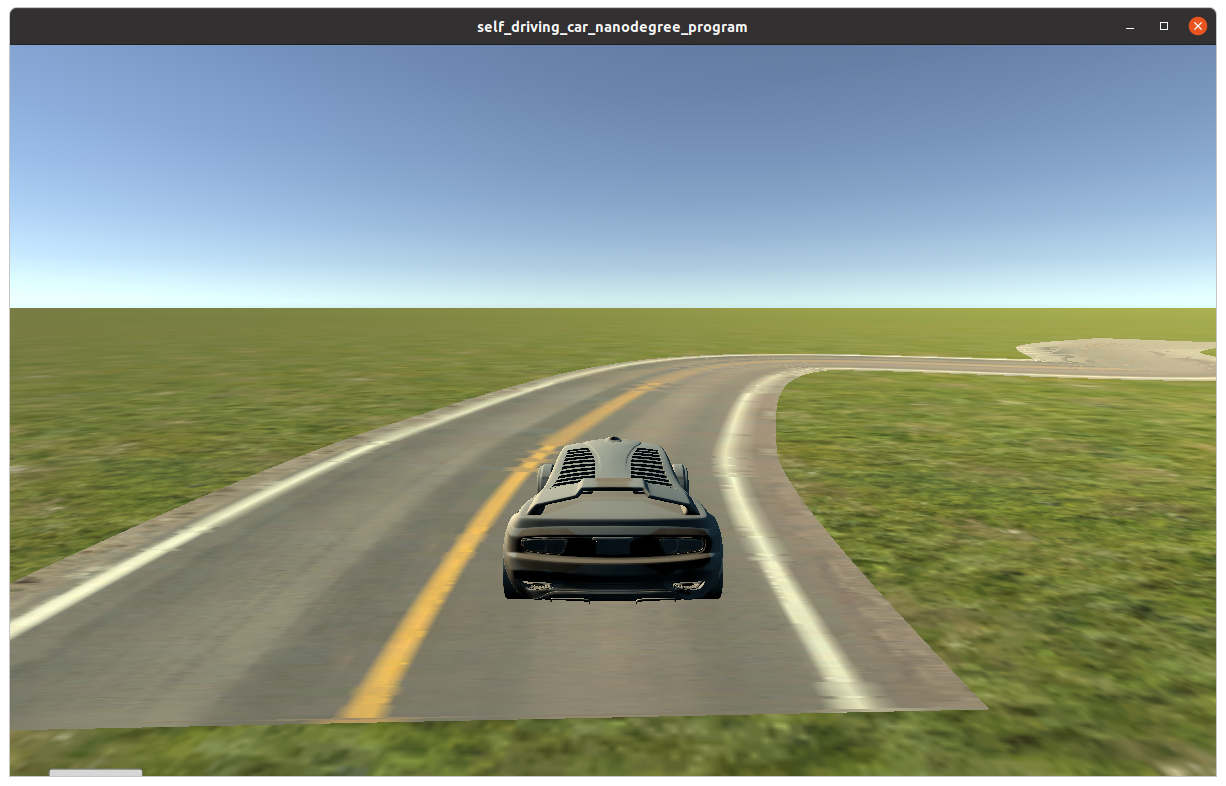}
        \caption{Udacity.}
    \end{subfigure}
    \hfill
      \begin{subfigure}[h]{0.3\columnwidth}
        \centering
      \includegraphics[trim={0.5cm 0cm 0cm 0cm},clip,scale=0.116]{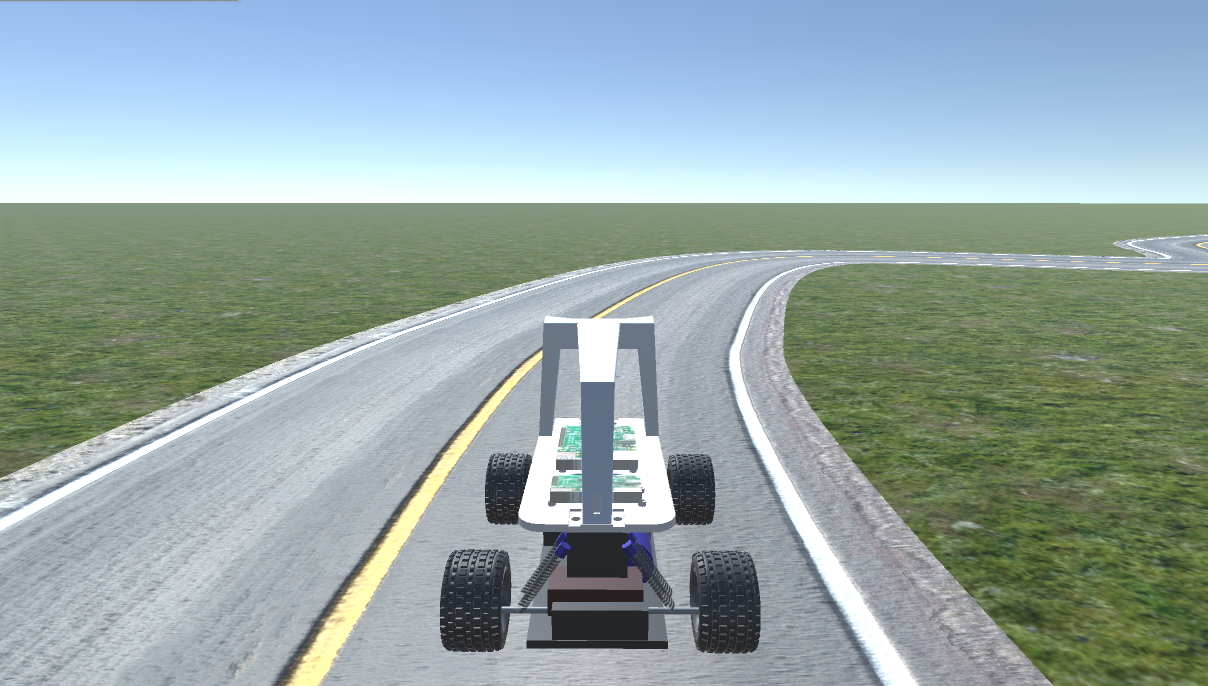}
        \caption{Donkey.}
    \end{subfigure}
    \hfill
      \begin{subfigure}[h]{0.3\columnwidth}
        \centering
        \includegraphics[scale=0.118]{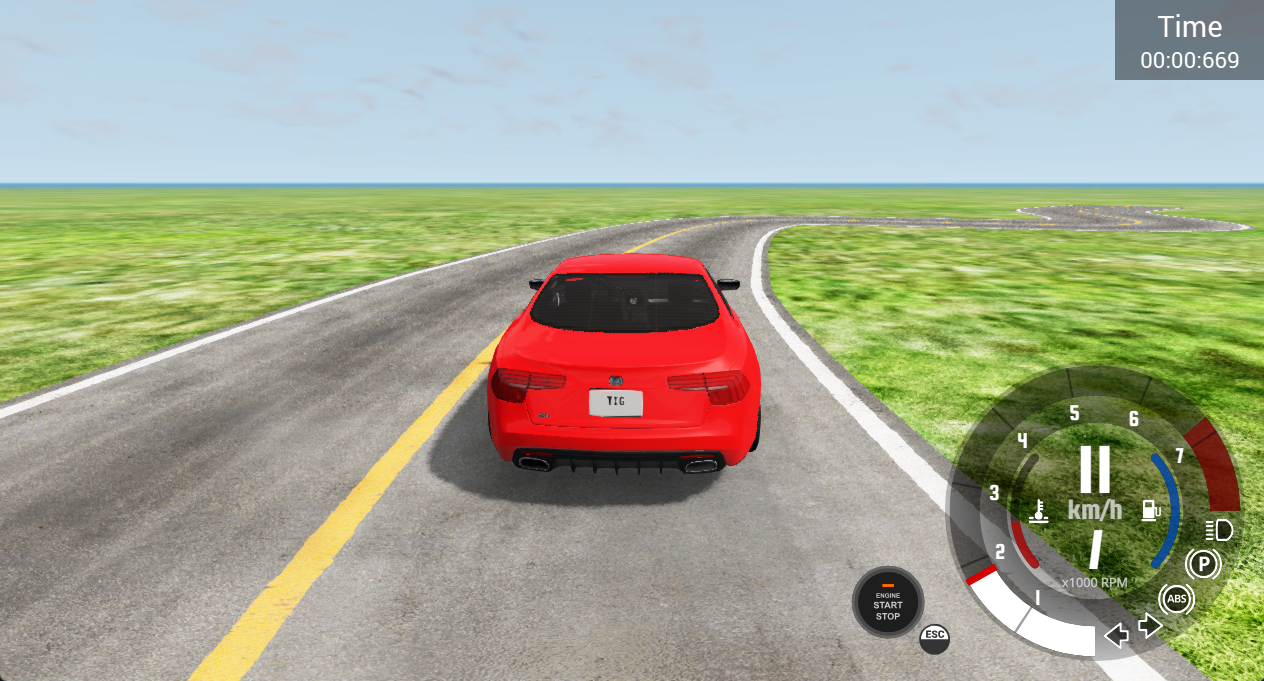}
        \caption{BeamNG.}
    \end{subfigure}

\caption{Difference in executing a lane-keeping ADAS on the same road in three different simulators, along with their rendering. In Udacity the trajectory of the vehicle (visualized in green, starting with a triangle) is within the lane's bounds while in Donkey and BeamNG the vehicle is departing off lane.}
  \label{fig:simulator_comparison}
\end{figure}

\subsection{Non-determinism in Virtual Testing}\label{sec:nondeterminism}

Recent studies have shown that there are several of simulation platforms available for system-level testing, both commercially and open-source~\cite{Li_2024,10174078}, with no consolidated omni-comprehensive solution. As multiple simulation platforms exist, researchers have started performing cross-replication studies of ADAS in different simulation platforms~\cite{AminiFlaky2024,Borg21CrossSimTesting}, to confirm the testing results obtained on a driving simulator on another, possibly independent, simulator. Ideally, when we execute the test input on the same ADAS in the different simulators, the pass/fail verdict should not change, providing trustworthiness to simulation-based testing.
However, the results of these studies highlight the negative aspects of simulation-based testing for ADAS. In most cases, the results obtained in one simulator (e.g., the failure conditions) cannot be reproduced by another simulator~\cite{AminiFlaky2024,Borg21CrossSimTesting,matteoMaxibon2024}. 
These discrepancies can lead to a distrust in simulation-based testing, as reported by recent surveys~\cite{icst-survey-robotics,fse-survey-robotics}. 

An example is shown in \autoref{fig:simulator_comparison}, in which we replicated the execution of a logical test scenario for a lane-keeping ADAS model on three simulators, namely Udacity~\cite{udacity-simulator}, Donkey~\cite{donkey} and BeamNG~\cite{beamng}. The logical test scenario includes a two-lane 70 meters long road with three curves, one on the right with curvature 0.24, followed by one on the left with curvature 0.22, and one on the right with curvature 0.08. The objective for the ADAS model is to keep the car within the right lane from the start of the road until the end. The figure depicts only the right lane and shows that the driving behavior of the lane-keeping ADAS model (shown by the green line trajectory) is quite different across simulators. While the car reaches its destination in all three examples, it yields different simulation traces, specifically failing in the Donkey and BeamNG simulators but not in Udacity. 

Potential root causes are related to flakiness in the simulation environments, bugs in the simulator, or synchronization issues in the communication between the simulator and the testing framework~\cite{Li_2024,10174078,Borg21CrossSimTesting,AminiFlaky2024}. 
As the simulation platform is used as-is, as a runtime testbed, it is usually not possible for developers to debug such issues, even less try to fix the simulator. 
This motivates the development of approaches to increase the reliability of simulation-based testing of ADAS, without requiring access to the simulation environments or its code, which is treated it as a black-box. 

\subsection{Simulator-agnostic Failures}

In this paper, we focus on identifying test scenarios where failures are not caused by the limitations of the simulation environment but rather by deficiencies in the system under test. To achieve this, we introduce the concept of \textit{simulator-agnostic failures}.  

Let us consider an ADAS $\mathcal{A}$ and a failing test input $\mathbf{x}$ that can be executed in a simulation environment $\mathcal{S}$. We define $\mathbf{x}$ as \textit{simulator-agnostic} failure-inducing test input if $\mathcal{A}$ also fails on $\mathbf{x}$ when executed in a different simulator $\mathcal{S}'$ with equivalent capabilities. In other words, the execution outcome of $\mathbf{x}$ should not depend on compatibility issues between $\mathcal{A}$ and the simulators, i.e., $\mathcal{S}$ and $\mathcal{S}'$. This concept can naturally be extended to multiple simulators. In this study, we consider three simulation environments and assume that if a failure is consistently observed across all three, it indicates a true defect in the ADAS, as it is independent of any single simulator. Conversely, failures that occur only in specific simulators are regarded as \textit{simulator-dependent}.  

Following the established terminology from the deep learning testing domain~\cite{2023-Riccio-ICSE}, we refer to simulator-agnostic failures as \textit{valid} failures, while we refer to simulator-dependent failures as \textit{invalid} failures throughout this paper.  

\section{Approach}
\label{sec:approach}

\subsection{Ensemble-based Test Case Generation}\label{sec:test-generation}

Our proposed \msim approach incorporates the evaluation capabilities of multiple simulators to identify simulator-agnostic failures.
The key idea is to apply ensemble learning within search-based ADAS testing, harnessing the strengths and compatibilities of individual simulators within a unified processing unit. 
\changed{We reformulated the multi-objective optimization problem to incorporate simulator-specific agreement and disagreement objectives, allowing the search to explicitly reason about the consistency of failures across heterogeneous simulators.}
Instead of evaluating a test input in one simulator, \msim combines the test input evaluation results of multiple simulators in one evaluation vector. 
Our approach aims to obtain more reliable, and hence more valid, failure-inducing test inputs than a single-simulator approach that only evaluates a test input in a single simulator.

\begin{algorithm}[t]
\DontPrintSemicolon
\footnotesize
    \SetKwInOut{Input}{Input}
    \SetKwInOut{Output}{Output}
    \Input{
        $(T, D, F, O)$: An SBST Problem. \newline
        $S = \lbrace S_1, \ldots, S_j \rbrace $: A set of simulators, with $j > 1$.
        
    }
    \Output{ $C$: Set of all failure-inducing test inputs.}
    \BlankLine
    \textsc{init}(T); $C \gets \emptyset$; $A \gets \emptyset$ \;
    $P_0 \gets \textsc{randomSampling}(T, D)$ \;
    $P_0 \gets \textsc{evaluateMsim}(P_0, S, F, T, A)$ \tcc*{\autoref{algo:eval-multisim}.}
    \While{$\textit{budgetAvailable}$}{
        $\textit{Par}$ $\gets$ \textsc{tournamentSelection}$(P_i)$\;
        $P_i$ $\gets$ \textsc{crossoverMutate}$(\textit{Par})$ \;
        $P_i$ $\gets$ \textsc{evaluateMsim}$(P_i,S,F,T,A)$ \tcc*{\autoref{algo:eval-multisim}.}
        \For{$\mathbf{x} \in P_i$}{
            $T \gets T \cup \lbrace \mathbf{x} \rbrace $ \tcc*{Store all evaluated tests.}
        }
        $P_i$ $\gets$ \textsc{survival}$(P_i,O)$\; 
        $P_i$ $\gets$ \textsc{repopulate}$(P_i,D,T)$
    }
    \For{$\mathbf{x} \in T$}{   
        $\textit{allFail} \gets \bigwedge_{k=1}^{j} O(f_k(\mathbf{x}))$
        \;
        \If{\textit{allFail}}{
        $C \gets C \cup \lbrace \mathbf{x} \rbrace$
        }
    }
    \KwRet{$C$}
\caption{Test case generation in \msim.}\label{algo:all-multisim}
\end{algorithm}

\autoref{algo:all-multisim} details the steps of the \msim approach.
The first step is to initialize the system under test $T$, the set of all failure-inducing test inputs $C$ as well as the archive $A$ (Line~1). Then, a random set of test inputs is sampled within the defined search domain (Line~2, $P_0$ being the initial population). In the next step, test inputs are evaluated based on the \msim specific evaluation function (see \autoref{algo:eval-multisim}). 
The main loop of the algorithm begins by applying default genetic operators such as selection, crossover and mutation (Lines~5--6). 
After the operators have been applied, a set of candidate test inputs to form the next population is defined.
In particular, when all newly generated test inputs have been evaluated in the \msim specific evaluation function (Line~3), the survival operator is applied to rank the test inputs based on dominance (Line~10). Then, the re-population operator is used to replace a portion of dominated tests of each population by randomly generated ones to diversify the search (Line~11). When the search budget is exhausted, each test input is marked as failing, when all simulators agree on the failure of such test (Line~12--15). Practically, the oracle function $O$ returns 1 for each simulator-specific fitness function $f_k(\mathbf{x})$, where $k: (i,\dots,j) < m$ indicates the fitness function for each specific simulator (Line~13).

\subsubsection{Multi-Simulator Evaluation}\label{sec:multisim-eval}

As detailed in \autoref{algo:eval-multisim}, in \msim the evaluation of a test input is defined as the fitness vector:

\begin{align}
   F(\mathbf{x}) = (f_1(\mathbf{x}), \ldots, f_j(\mathbf{x}), f_d(\mathbf{x}), f_a(\mathbf{x}, A))
\end{align}

\noindent
where (1)~$f_1(\mathbf{x}) = v_1, \ldots, f_j(\mathbf{x}) = v_j$ represent the fitness functions for the execution of $\mathbf{x}$ in simulator $S_1, \ldots, S_j$ (Lines~3--4), (2)~$f_d(\mathbf{x}) = v_d$ is defined as the average distance between the fitness values $f_1(\mathbf{x}), \ldots,  f_j(\mathbf{x})$ (Line~5), and (3)~$f_a(\mathbf{x}, A) = v_a$ is the distance of the test input $\mathbf{x}$ to previously found test inputs in the archive $A$ (Line~6).

\begin{algorithm}[t]
\DontPrintSemicolon
\footnotesize
    \SetKwInOut{Input}{Input}
    \SetKwInOut{Output}{Output}
    \Input{
        $T:$ System under test. \newline
        $S = \lbrace S_1, \ldots, S_j \rbrace $: A set of simulators, $j > 1$. \newline
        $\delta$: distance threshold \newline
        $P$: Not evaluated test inputs. \newline
        $A$: Archive of test inputs.
    }
    \Output{$Q$: Evaluated test inputs. }
    $Q \gets \emptyset$\;
    \BlankLine
    \For{$\mathbf{x} \in P_i$}{
        \For{$S_k \in S$}{
            $v_{k} \gets \textsc{simulate}(\mathbf{x}, T, S_k)$ \tcc*{Evaluate fitness for each simulator.}
        }
        \blue{$v_d \gets \frac{\sum\limits_{k=1}^{j-1} \sum\limits_{l=k+1}^{j} |v_k - v_l|}{\binom{j}{2}}$ \tcc*{Compute average distance between fitness values.}}
        $v_a \gets \textsc{getDistanceToArchive}(\mathbf{x},A)$ \tcc*{Compute distance w.r.t. archived test inputs.}
        $\mathbf{x}.F \gets (v_1, \ldots, v_j, v_d, v_a)$ \tcc*{Assign fitness values to test input.} 
        $\textsc{addToArchive}(\mathbf{x},A,\delta)$ \; 
        $Q \gets Q \cup \lbrace x \rbrace$
    }
    \KwRet{$Q$}
\caption{Test input evaluation in \msim.}\label{algo:eval-multisim}
\end{algorithm}

While optimizing the functions $f_1(\mathbf{x}), \ldots, f_j(\mathbf{x})$ prioritizes failure-revealing test inputs, minimizing $f_d(\mathbf{x})$ drives the search towards test inputs on which all simulators agree. The fitness function $f_a(\mathbf{x}, A)$ aims to diversify the search by calculating the distance of a test input to previously generated tests in the archive $A$~\cite{2020-Riccio-FSE}. As distance metric we use the Euclidean distance between normalized test inputs. In particular, we apply Min-Max normalization, scaling values in the test input representation to a range from 0 to 1.
By maximizing $f_a(\mathbf{x}, A)$ the goal is to identify more diverse test cases. After the evaluation of a test input is completed, it is added to the archive in case its distance to the closest test in the archive exceeds a predefined threshold $\delta$ (the function \textsc{addToArchive} in Line~8, takes care of comparing the current test input $\mathbf{x}$ with those in the archive $A$). 

\subsubsection{Test Representation}\label{sec:representation}

\begin{figure}[t]
       \raggedleft
    \centering
    \begin{subfigure}[h]{0.35\columnwidth}
        \centering
        \includegraphics[scale=0.25]{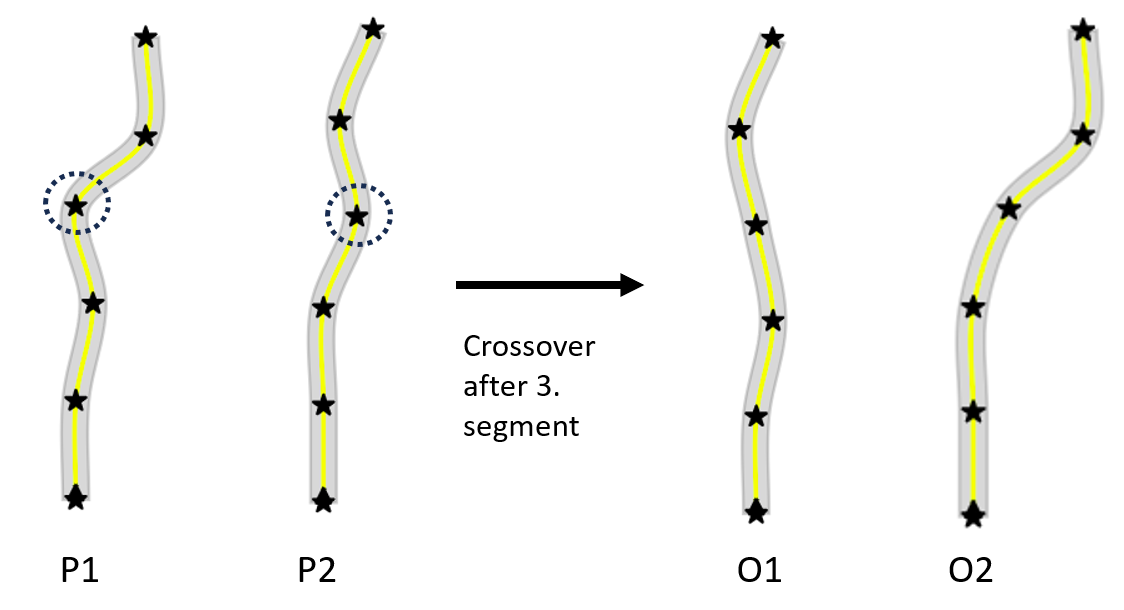}
        \caption{Crossover.}
        \label{fig:crossover}
    \end{subfigure}
    \hfill
      \begin{subfigure}[h]{0.35\columnwidth}
        \centering
        \includegraphics[scale=0.25]{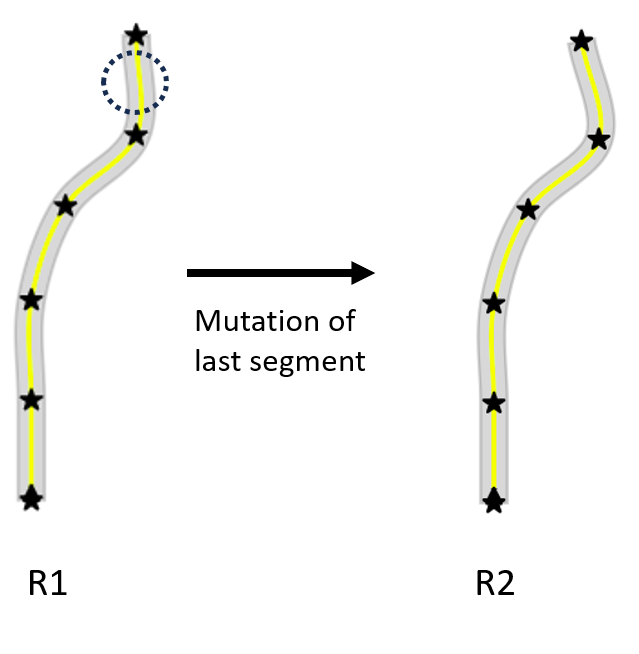}
        \caption{Mutation.}
        \label{fig:mutation}
    \end{subfigure}
    \caption{Illustration of crossover (a) and mutation (b) of roads in \msim including control points which are placed within the road (visualized as stars). For sake of simplicity only angles are modified. In Figure a) tails of roads after the third segment are exchanged. In Figure b) the angle of the last segment is increased by 10 degrees.}
    \label{fig:mutation-crossover}
\end{figure}

\changed{We adopt a model-based representation of test inputs. In line with previous studies~\cite{2020-Riccio-FSE,2025-Ali-ICSEW}, each road scenario is defined through a sequence of control points. These points are interpolated using Catmull–Rom splines to generate smooth and continuous trajectories.}
To be able to generate diverse and challenging roads we parameterize the road definition using the vector $\mathbf{x} = (\alpha_1, \ldots, \alpha_n, l_1, \ldots, l_n)$, where $a_i \in [-180; 180)$ denotes the clockwise angle between the i-th segment and the horizontal axis, while $l_i \in \mathbb{R}^+$ represents the length of the i-th segment between the control points $c_{i+1}$ and $c_{i+2}$. For instance, the road in \autoref{fig:crossover} has 7 control points, 5 segments and is represented by the vector~$(-90, -110, 173, -120, -140, 10, 20, 18, 17, 15)$. 
\changed{Note that this choice of input representation is not exclusive of our approach and equivalent parameterized representations can be defined, following similar principles~\cite{2025-Chen-arxiv}.}

\subsubsection{Mutation/Crossover}

Our implementation of the crossover operator on roads follows that of existing studies~\cite{Humeniuk24Reinforcement,moghadam2023machine}. Specifically, we use a one-point crossover operator, where for two given roads $\mathbf{p}_1$, $\mathbf{p}_2$ (i.e., the parents), we first randomly select an index, followed by exchanging the tail of the road $\mathbf{p}_2$, including segments with a lower index, with the tail of road $\mathbf{p}_1$, including segments with a higher index. \autoref{fig:crossover} illustrates the crossover of the road $\mathbf{p}_1$ with the road $\mathbf{p}_2$ resulting in the roads $\mathbf{o}_1$ and $\mathbf{o}_2$ (i.e., the offsprings), by exchanging the tails of the roads after the third segment.

The mutation of a road $\mathbf{x}$ is performed by randomly selecting a segment, followed by increasing or decreasing the angle, or the segment length, by a given amount. A mutated road can be invalid because of intersecting segments being placed out of the given map, or when violating a threshold which defines the maximum angle between segments. If a road is invalid, we randomly generate a new road by applying the same operator as used to generate the initial population for \msim.
As an example, \autoref{fig:mutation} shows the mutation of a road $\mathbf{r}_1$ by increasing the angle of the last segment from 90$^{\circ}$ to 100$^{\circ}$.

\subsubsection{Fitness Function}

\changed{
For the evaluation of a single test input within one simulator, we define a fitness function that captures the system-level behavior of the SUT. In the case of lane-keeping ADAS, the chosen fitness function is the cross-track error, denoted as $f_{\textit{XTE}}$. This function computes the maximum lateral deviation between the vehicle's center and the lane center over the entire duration of the test execution, providing a direct quantitative measure of lane-keeping performance.} 
We selected this function because it has been shown to be an effective fitness function for triggering failure-inducing test inputs~\cite{matteoMaxibon2024, Humeniuk24Reinforcement, 2020-Riccio-FSE}. As oracle function we use $O = f_{\textit{XTE}} > d$ which evaluates a scenario execution as a failure when the maximal \textit{XTE} value exceeds $d$.

\subsection{Test Validation}\label{sec:test-validation}

\begin{figure}[t]
\centering
    \includegraphics[width=0.6\linewidth]{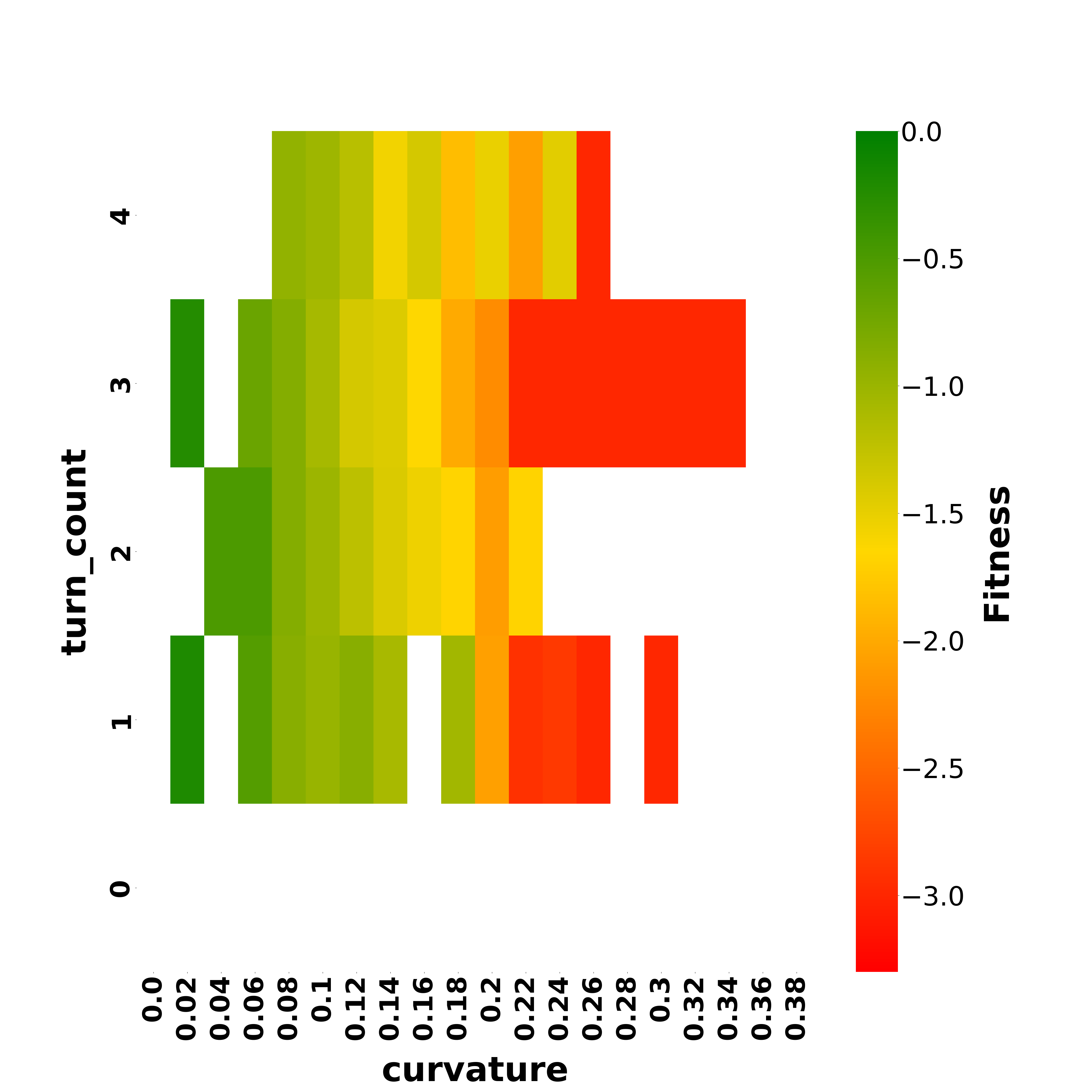}
\caption{Example of a feature map with two features, namely number of turns (i.e., \texttt{turn\_count}) and curvature. The color of a cell is defined based on the worst \textit{XTE} value of test inputs stored in a cell (colorbar on the right-hand side of the map). A cell is green when the \textit{XTE} value is 0, red for the maximum, in absolute value, \textit{XTE} value of -3. A cell is white if there is no test input that \textit{covers} it. The fitness is negative, because the fitness function is to be minimized. In total 412 tests are stored in the feature map, which has 13 failing-cells and 35 non-failing cells.}
\label{fig:feature-maps}
\end{figure}

This step aims to validate that the identified failing tests are not failing due to simulator-specific behaviors. To characterize the ADAS unique failures, we adopt the feature maps by Zohdinasab et al.~\cite{zohdinasab2021deephyperion, mapelites}.

\head{Feature Maps} In the first step of our validation approach, we map test inputs generated by \msim into a feature space~\cite{zohdinasab2021deephyperion, mapelites}, as it allows to represent important characteristics of test inputs in a human-interpretable way. In particular, with a feature space we can discretize the multi-dimensional search space, which is the space of roads (see \autoref{sec:representation}), for our lane-keeping ADAS, and consider diverse test cases for the evaluation of our approach. Our feature space is characterized by two static road attributes, i.e., \textit{num of turns} and \textit{curvature}, i.e., the reciprocal of the minimal value of the radius over all circles that can be placed through three consecutive control points of the road. Similarly to the fitness function, such choices of features were also found to be effective at characterizing the search space of road generators~\cite{zohdinasab2021deephyperion}.

An example of a feature map is shown in \autoref{fig:feature-maps}. 
A cell with the coordinates $(x,y)$ contains test inputs whose curvature is in the range $[x - 0.02,\ x + 0.02]$, where $0.02$ is the granularity of the map, and the number of turns is $y$.
For instance, the road in \autoref{fig:flakiness_low} is to be placed in the cell $(0.24, 3)$ as it has a maximum of curvature 0.23 and three turns. A cell of a feature map is colored based on the road with the best fitness placed in the cells, which is ranging from green (worst, $f_{\textit{XTE}} = 0 $) to red (best, $f_{\textit{XTE}} = -3$) fitness value. We deem a cell as \textit{failing} cell, when at least one test in the cell is failing w.r.t. the given oracle $O$.
For instance, the cell $(0.32, 3)$ is a failing cell, while the cell at the position $(0.04, 2)$ is non-failing cell. 

\begin{figure}[t] 
    \centering
    \includegraphics[scale=0.6]{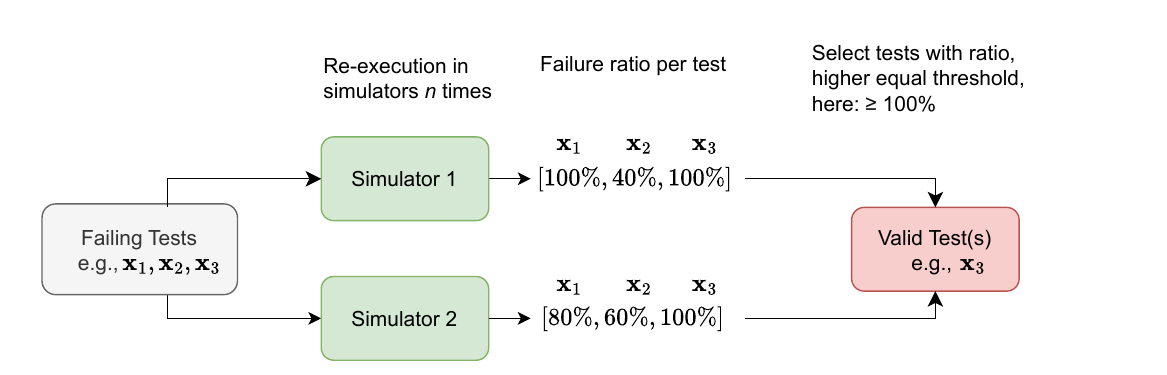}
    \caption{Validation of failure-inducing test inputs, i.e., $\mathbf{x}_1$, $\mathbf{x}_2$, $\mathbf{x}_3$. Each test is re-executed in multiple simulators, i.e., two in this case $S_1$ and $S_2$, and the failure rates for each test are computed. Then the failure rates on $S_1$ and $S_2$ are compared, and the test inputs are filtered according to a failure rate threshold. (in this case 100\%). In this example, the only simulator-agnostic/valid failure-inducing test input is $\mathbf{x}_3$. }
    \label{fig:msim-validation}
\end{figure}

\head{Re-execution} In the second step, we select failing cells from the feature map, extract a portion of failing and duplicate-free test inputs per cell, and re-execute each of these selected failure-inducing test inputs multiple times, to account for the flakiness of each specific simulation environment. 
Specifically, we execute such test inputs on simulators that have been not used by \msim during its search process, to confirm whether the failure-inducing test inputs generated by \msim are simulator-agnostic or not. For the validation we consider the notion of \textit{hard} flakiness as defined by Amini et al.~\cite{AminiFlaky2024}, where a test input is considered hard flaky when re-executing the test yields different pass or fail verdicts.  
\autoref{fig:msim-validation} illustrates the validation approach, when employing two simulators for the validation. First, each failure-inducing test input is executed on each simulator for $n$ times. For each execution the test input is evaluated based on the fitness and oracle functions (the same used during the test case generation process of \msim). Afterwards, a failure rate is computed.
Finally, the failure rates are compared between the simulators for each test input to decide if the test is failing according to a user-defined threshold. For instance, in \autoref{fig:msim-validation} we have selected a 100\% threshold, i.e., a failure-inducing test input is considered as simulator-agnostic if for every re-execution and for every simulator the test fails.
In the example in \autoref{fig:msim-validation} only the test $\mathbf{x_3}$ is a simulator-agnostic/valid failure-inducing test input, as both simulators, i.e., simulator $S_1$ and $S_2$, report a 100\% failure rate. For $\mathbf{x_1}$ a re-execution in $S_1$ yields a failure rate below 100\%. Similarly, for $\mathbf{x}_2$ re-executions in both simulators $S_1$ and $S_2$, achieve failure rates below the threshold.

\section{Empirical Study}\label{sec:study}

In this section, we present the evaluation of our approach for identifying simulator-agnostic failures for a lane-keeping ADAS case study. We describe the research questions, our evaluation technique, evaluation experiments, implementation and results.

\subsection{Research Questions}
\label{sec:rqs}
We consider the following research questions:

\noindent\textbf{RQ\textsubscript{1} (effectiveness).} \textit{How effective is \msim in generating simulator-agnostic failure-inducing test inputs compared to single-simulator testing and a state-of-the-art approach?}

The first question evaluates the effectiveness of \msim in finding simulator-agnostic failures. After the test case generation phase, we validate a portion of failure-inducing test scenarios based on the approach described in \autoref{sec:test-validation}. 

\noindent\textbf{RQ\textsubscript{2} (efficiency).} \textit{How efficient is \msim in generating simulator-agnostic failures compared to single-simulator testing and a state-of-the-art approach?}

The second research question evaluates whether \msim is efficient in identifying failures. 
Test efficiency is particularly relevant when the testing budget is limited~\cite{Raja16NeuralNSGA2}. 

\noindent\textbf{RQ\textsubscript{3} (prediction).} \textit{To what extent can machine learning predictors increase the effectiveness and the efficiency of \msim in finding simulator-agnostic failures?}

Since simulator-specific failures, i.e., disagreements between simulators, might impact the identification of simulator-agnostic failures during the search, we use machine learning (ML) classifiers to predict whether a test input is leading to a disagreement between pass/fail verdicts on different simulators.
We investigate whether this approach can be considered as a \text{calibration} technique to fine-tune the original \msim approach to mitigate disagreements during the search caused by simulator-specific failures. 

\subsection{Baselines}\label{sec:baselines}

We compare the performance of \msim to a search-based testing approach which only employs one simulator (\ssim) during the search. 
We also compare with the recently proposed \dss approach~\cite{matteoMaxibon2024}, which employs multiple simulators by re-executing test inputs generated for one simulator in a different simulator and vice-versa. \dss is, to the best of our knowledge, the only existing approach in the literature that employs multiple simulators for automated test case generation in the ADAS domain.

\subsection{Procedure and Metrics}\label{sec:metrics}

To answer all three research questions for our experiment, we first validate the failures as described in \autoref{sec:test-validation}. Specifically, we assign all tests to cells of the feature map, by extracting curvature and number of turns of the corresponding roads, using the same feature space granularity as proposed by Biagiola et al.~\cite{matteoMaxibon2024}. 
We randomly select three failure-inducing test inputs from each failing cell. Preliminary experiments with a higher number of selected tests per failing cell, i.e., five, did not affect the results significantly. 

To answer RQ\textsubscript{1}, to validate failures found by \msim, we use the third simulator that was not used during the search. We then re-execute each test five times to account for the flakiness of the simulation environment. As failure rate threshold we use 100\%, i.e., a failure-inducing test input found by \msim is only valid if it always fails in the third validation simulator. 

To validate failures found by \ssim, we use the same validation approach, but we use the two simulators that are not used during the search. For \dss we also use only one simulator for validation, as two of the three simulators available for our study are already used in the search. Based on the number of selected failures and the validation results, we assess the performance of \msim, \ssim and \dss using two metrics: (1) \textit{n\_valid}, which is the number of valid failures, and (2) \textit{valid\_rate}, which is defined as:

\begin{align*}
    \text{valid\_rate} = \frac{\text{\# valid failures}}{\text{\# failures selected for validation}}
\end{align*}

To answer RQ\textsubscript{2}, after validating all the selected failures, we identify at which time during the search each valid failure is encountered. 
\changed{Thus, we define efficiency as the proportion of the search budget required to identify the first valid failure. It measures how early in the search process the approach discovers a fault-revealing test relative to the total number of generations or evaluations. This metric reflects the practical usefulness of the search procedure in time- and resource-constrained testing contexts, where faster identification of valid failures is beneficial.}

For \dss, as it consists of two independent search executions and two migration steps, we first identify in which simulation run the failure is detected. Based on the order of the run executions and the evaluation number we compute the total run time to find the failure. That is, when a failure is detected in simulator DS1 with in total $A$ test evaluations at the evaluation number $a$, while simulator DS2 performed in total B test evaluations, we assign to the test the final evaluation count as $A + B + a$. 
In case the valid failure is detected during the execution of $B$ at evaluation number $b$, we assign to the test the evaluation count $A + B + A + b$, respectively.
\blue{We evaluate efficiency using this methodology because DSS consists of two independent testing experiments, and unlike single-simulator-based testing, it is not possible to incrementally order the test cases generated by DSS for a single run.}

To answer RQ\textsubscript{3}, we select the configuration of \msim with the \textit{best} results in terms of validity rate and number of failures, and collect test cases from past executions where the simulators both agree or disagree. In addition, we execute a modified version of \msim with five different seeds where the fitness function prioritizes disagreements. Specifically, we reuse \autoref{algo:eval-multisim}, but maximize $f_d(\mathbf{x})$, i.e., the distance between maximum XTE evaluations. 
We pass the roads as inputs and the labels representing an agreement or disagreement to five different and widely used ML classification models for training. Specifically, we use Decision Trees~\cite{breiman1984classification}, Random Forests~\cite{breimannRandomF2001}, Support Vector Machines~\cite{cortesSvm95}, Logistic Regression~\cite{cabrera1994logistic} and Stochastic Gradient Boosting~\cite{FRIEDMAN2002367}. For the training of the classification models we use default hyperparameter configurations available in the Sklearn library. 

We then evaluate the ML models based on F-1-score and AUC-score. 
The best performing ML model on all metrics is then integrated into \msim and executed with the same search configuration as \msim.

In terms of metrics, we measure the number of valid failures (\textit{n\_valid}) and validity rate (\textit{valid\_rate}) as in RQ\textsubscript{1}, and compare \msim with and without the integration of a disagreement predictor. We also measure the first valid failure as the efficiency metric for both configurations.

\subsection{Experimental Setup}\label{sec:setup}

\subsubsection{Simulation Environments} 

We have selected the following simulators for our study: BeamNG, Udacity and Donkey Car.
\changed{
We have chosen such simulators for the following reasons: 1) they seamlessly integrate with our lane-keeping ADAS case study, as the simulation environment, i.e., asphalt road on green grass and sunny weather, has been aligned~\cite{matteoMaxibon2024} to be able to replicate the same simulation conditions across different simulators (see \autoref{fig:simulator_comparison});} 2) they are complementary as they differ in terms of the physics implementation or the rendering engine; 3) they are open-source or available under permissive academic-licenses (BeamNG). A brief description of the simulators provided is as follows:

\begin{itemize}
    \item BeamNG is based on Unreal engine, implements soft-body physics, serves as the simulator in the SBFT competition~\cite{gambi22sbft, liu23sbft, matteo24sbft}, and it has been used extensively in the literature to benchmark testing approaches~\cite{beamng, zohdinasab2021deephyperion, 2020-Riccio-FSE, Humeniuk24Reinforcement}. 
    \item The Udacity simulator is a cross-platform and rigid-body physics developed with Unity3D~\cite{udacity-simulator}. While originally developed for educational purposes, it has been widely applied for ADAS testing~\cite{matteoMaxibon2024, 2024-Grewal-ICST,2022-Stocco-ASE,2021-Stocco-JSEP,2021-Jahangirova-ICST,DeepGuard,2025-Baresi-ICSE}.
    \item The Donkey Car simulator (Donkey, hereafter) is part of the Donkey Car framework that is used to build and test small-scale self-driving physical vehicles~\cite{donkeycar}. Similarly to Udacity, it implements rigid-body physics, and is developed with Unity 3D. The simulator is actively used in the literature for the validation of testing approaches~\cite{matteoMaxibon2024,2023-Stocco-TSE,2023-Stocco-EMSE,survey-small-scale-2,survey-small-scale-1,2025-Lambertenghi-ASE}.
\end{itemize} 

\changed{Unlike previous work~\cite{matteoMaxibon2024}, we make no assumptions about simulator fidelity and treat all simulators as black-box environments.}
In our experiments, we execute \msim in three different configurations: in the configuration BD we use BeamNG and Donkey as simulator ensemble, in the configuration BU we use BeamNG and Udacity, and lastly in the configuration UD we use Udacity and Donkey. We also evaluate \dss in such configurations; in the tables and figures, we prefix such configurations with \dss- (e.g., \dss-BU indicates the configuration BU with the \dss approach, while BU indicates the same configuration with the \msim approach). It is noteworthy that in the original DSS paper~\cite{matteoMaxibon2024} only the configuration \dss-BU was proposed and evaluated. The extension of the approach to other simulators constitutes a novel experimental contribution of our work.

\subsubsection{Study Subjects} 

\changed{To assess the generality and robustness of our approach, we evaluate it on three lane-keeping ADAS that differ in their design paradigms, from convolutional models, to transformer-based architectures leveraging global attention, and hybrid vision–trajectory policies that integrate planning information, described next:}

\begin{itemize}

\item \textbf{CNN-based ADAS (\davetwo)}: \davetwo is a convolutional neural network developed for multi-output regression tasks based on imitation learning~\cite{nvidia-dave2}. The model architecture includes three convolutional layers for feature extraction, followed by five fully connected layers. The model take as input an image representing a road scene, and it is trained to predict vehicle's actuators commands. We use a pre-trained \davetwo model from previous work~\cite{matteoMaxibon2024} that has been trained on more than 60,000 images and steering commands automatically by a PID-controlled autopilot driving the car on roads with different road complexity in the three simulators used in this work.
\davetwo has been extensively used in a variety of ADAS testing studies~\cite{2023-Stocco-TSE,deeptest,10.1145/3238147.3238187,matteoMaxibon2024,2021-Jahangirova-ICST,biagiola2023boundary,2024-Lambertenghi-ICST,2025-Lambertenghi-ICST}. 

\item \blue{\textbf{Vision Transformer-based ADAS (\vit)}: \vit~\cite{vit} is a DNN based on the transformer architecture to predict steering commands. While Transformers were originally developed for natural language generation and understanding, they have since been widely adopted for visual tasks~\cite{RangeViT}, including lane-keeping, where \vit achieves results competitive with systems such as \davetwo~\cite{2025-Baresi-ICSE}. \vit receives an image of the road scene and predicts the steering command.
Similarly to \davetwo, we trained the \vit model using images and steering commands collected by the PID controller driving the car on 100 tracks for each simulator. Differently from \davetwo, we trained ViT using approximately twice as many images (i.e., 120,000).}

\item \blue{\textbf{Trajectory-Control Policy ADAS (\ticp):} \ticp~\cite{tcp} is a hybrid deep neural network that combines vision-based perception with trajectory-conditioned control to predict steering and throttle commands. The model adopts a two-branch architecture: a CNN-based vision encoder that extracts high-level spatial features from the input image, and a trajectory encoder (an MLP) that processes the planned waypoints provided by the controller. The two representations are fused in a joint latent space and passed through fully connected layers to produce continuous control outputs.
Training follows a procedure similar to \davetwo and \vit. We collected synchronized driving data from 40 tracks per environment by running a PID controller to generate reference control signals. Each sample includes the RGB scene image, the corresponding trajectory points, and the ground-truth steering and throttle commands. The final dataset comprises over 62,000 samples, enabling \ticp to learn a context-aware mapping between the visual scene, planned path, and control actions.}

\end{itemize} 

\subsubsection{Search Setup}

\head{Search Algorithm} 
For the test case generation with \msim we use \autoref{algo:all-multisim}.
For \ssim, we employ only one simulator for the evaluation of the fitness value of a test case. We use the same hyper-parameter configuration for all algorithms, such as mutation, crossover rate and population size. 
To select the archive threshold $\delta$ that controls which test inputs shall be included in the archive during search, we follow the guidelines of Riccio et al.~\cite{2020-Riccio-FSE} and perform preliminary experiments with different thresholds, ranging from 4 to 6. We assess the diversity of the corresponding feature map after the search, by identifying how many cells of the map are covered. As the archive threshold affects the diversity of the search, we use the coverage of the feature map to identify an appropriate threshold, which is 4.5 in our case. 
Further, we configure a maximum search budget of six hours for all testing approaches with \davetwo and \vit and 3.5 hours with \ticp, which proved sufficient to explore the search space and obtain an adequate number of failures.

In the \dss approach, we also use \autoref{algo:all-multisim} for test generation. Test cases are produced through two independent search executions, each employing a different simulator, i.e., $S_1$ and $S_2$. The test cases generated by $S_1$ are subsequently re-executed in $S_2$, and vice-versa. A test case is classified as critical only if it has been evaluated as critical by both simulators. The results from both simulators, including all generated and migrated test cases, are then aggregated to represent the final output.

\begin{table}[t]
    \centering
      \caption{Test generation configuration for the approaches under test \msim, \ssim and \dss. All configurations are equal across methods, beside the number of fitness functions used for \msim.}
    \label{tab:experiment_configuration}
    \begin{tabular}{lccc}
    \toprule
    \textbf{Parameter} & \textbf{\ssim} &  \textbf{\msim} & \textbf{\dss} \\
    \midrule
    Population size & 20 & 20 & 20\\
    Time budget & 6 hours & 6 hours & 6 hours\\
    Mutation rate & 1/10 & 1/10  & 1/10\\
    Crossover rate & 0.6 & 0.6 & 0.6\\
    Archive distance &  4.5  & 4.5  & 4.5\\ 
    \# Fitness functions &  2 & 4 & 2\\ 
    \# Search variables &  10 & 10 & 10\\
    \bottomrule
    \end{tabular}
\end{table}

For \dss, we adapted the search budget for the search in two selected simulators before the test migration: we set it to a quarter of the total search budget, because first, two independent search runs are executed, and in the second step results have to be migrated in the respective simulator. The remaining parameters are set as for \ssim and \msim.
An overview of the complete configuration for all search methods can be found in \autoref{tab:experiment_configuration}.

\head{Initial Population} To generate the initial population of roads, we use the test generator implemented in the study from Riccio et al.~\cite{2020-Riccio-FSE}, where a road is iteratively generated while persevering the user-defined constraints such as segment length and maximum angle between consecutive segments. In particular the algorithm samples segments randomly based on the given maximum angle and segment length.

\head{Mutation/Crossover} The mutation extent is randomly selected in the range of $[-8, 8]$, while the segment length is randomly set in the range $[10, 20]$. 

\head{Fitness/Oracle Function} As fitness and oracle functions we use the function $f_{\textit{XTE}}$ as defined in the motivation example. As oracle function we use $O = f_{\textit{XTE}} > 2.2$ that evaluates a scenario execution as a failure when the vehicle leaves the driving lane (similar to existing case studies~\cite{humeniuk2022searchbased, matteoMaxibon2024}). We stop the simulation on a given road when the current XTE value exceed a threshold of 3 mt, or the car reaches the end of the last segment. 

\subsection{ADAS Preliminary Evaluation}
\label{sec:prelim-experiments}

For \davetwo, we considered an ADAS model from previous work~\cite{matteoMaxibon2024}, to mitigate the threats to validity of training our own model for testing.  Vision perception models like \davetwo, are negatively affected by changes of image distributions~\cite{2024-Lambertenghi-ICST,2025-Lambertenghi-ICST,2025-Lambertenghi-ASE}. Thus, we assess the driving performance of the ADAS from previous work in our environment prior to the empirical evaluation. Specifically, we have defined 16 different types of roads with a road-wise maximum curvature ranging from 0 to 0.19, while the number of turns was in the range 1 to 4, and the road length was 100 mt. The evaluation allows us, on the one hand to investigate the flakiness of the simulation environments, and on the other hand to verify whether the testing setup including the hardware and software configurations is suitable for our study.

For \vit, we selected an existing model architecture~\cite{2025-Baresi-ICSE} and trained the model on collected images with corresponding PID-based steering commands by letting the model drive on 100 diverse automatically generated roads. 

\changed{For \ticp, we reused the model architecture from the original work~\cite{tcp}, which includes a ResNet model for visual feature extraction trained in the CARLA simulator. Because the pre-trained ADAS exhibited consistent failures across our simulators, we retrained it on a unified dataset collected from all three environments through PID-controlled driving over 40 randomly generated tracks, ensuring balanced exposure to the visual and physical characteristics of each simulator. The collected data was composed of images, steering commands, vehicle position information, and target waypoint positions. The TCP model defines a parameter \textit{alpha} that specifies the weight of the \textit{trajectory branch} vs. the \textit{predict control branch} for combined steering value prediction. We selected the alpha score, i.e., 0.1, based on preliminary experiments using validation roads.}
We executed all ADAS in every simulator on each of the predefined roads for 10 times to account for the flakiness, and assessed the quality of the driving using the XTE fitness function.

The results show that all ADAS exhibit low XTE variation (below 0.7m) across all roads without failing, even on roads with a high number of turns and high curvature. 
These observations suggest that while the vehicle's trajectories and driving styles are not harmonized across simulators, the ADAS's driving performance is not impacted by hardware limitations or latencies, as the variation is minimal and the ADAS does not fail. Thus, we conclude that the evaluated ADAS under test are sufficiently robust for exhaustive testing.

The results of all preliminary experiments  with a detailed description of the roads used can be found in our replication package~\cite{replication-package}.

\subsection{Implementation}
\label{sec:implementation}

We have implemented \msim using the testing framework OpenSBT~\cite{sorokin2023opensbt}. OpenSBT is a modular framework that eases the execution, as well as the extension of SBT components for conducting testing experiments. OpenSBT has been recently applied for evaluating testing techniques and replication studies~\cite{Sorokin2024, nejati2023reflections, sorokin24svm}. To implement our test generator in OpenSBT, we have extended the problem definition and test execution interfaces by evaluating a test input in multiple simulators and merging the result using a migration function.
Our implementation allows to use any simulator and evaluation migration strategy, which is integrated into the framework.  The code of \msim is available in our replication package~\cite{replication-package}.

All experiments have been executed on two computing devices, one PC with a Ryzen 5 5600X CPU, 32 GB RAM, and a 4090 Ti GPU, and on a Linux Laptop with an i7 CPU, 32 GB RAM, and a 790 GX GPU. To accelerate DNN computations we used CUDA 11.8 with cuDNN 8.9. Note, that due to simulator-specific requirements a considerable effort was required to identify a hardware configuration which was able to support our case study. BeamNG exhibited initially a very low FPS number when running on a device with the GPU 4090 Ti. In addition, BeamNG does not support the Linux, but runs on Windows. Udacity, in contrast, could not run on our Windows machine, so that we had to employ a second computing device, i.e., a Linux laptop, to perform the experiments for the BU/DSS-BU configuration. Finally, we made sure that all simulators run at 20 FPS. 

\changed{Overall, our experiments required approximately 1,395 hours i.e., $\approx$ 58 days (3 systems $\times$ 6/10 repetitions $\times$ 9 methods $\times$ 3.5/6 hours) for all compared test generation approaches. Additional more than 25 hours were required to validate more then 6,000 failing tests.}

\subsection{Results}

\subsubsection{Effectiveness (RQ\textsubscript{1})}

\begin{figure}[htbp]
    \centering
    
    \begin{subfigure}{0.85\linewidth}
        \includegraphics[width=\linewidth]{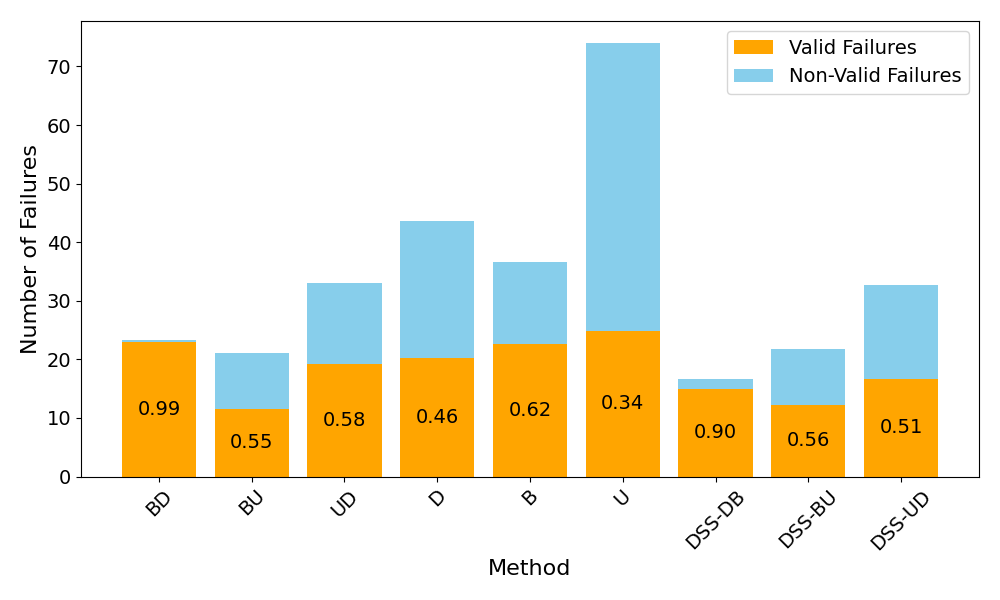}
        \caption{\davetwo}
    \end{subfigure}
    
    \vspace{0.5em}
    
    \begin{subfigure}{0.85\linewidth}
        \includegraphics[width=\linewidth]{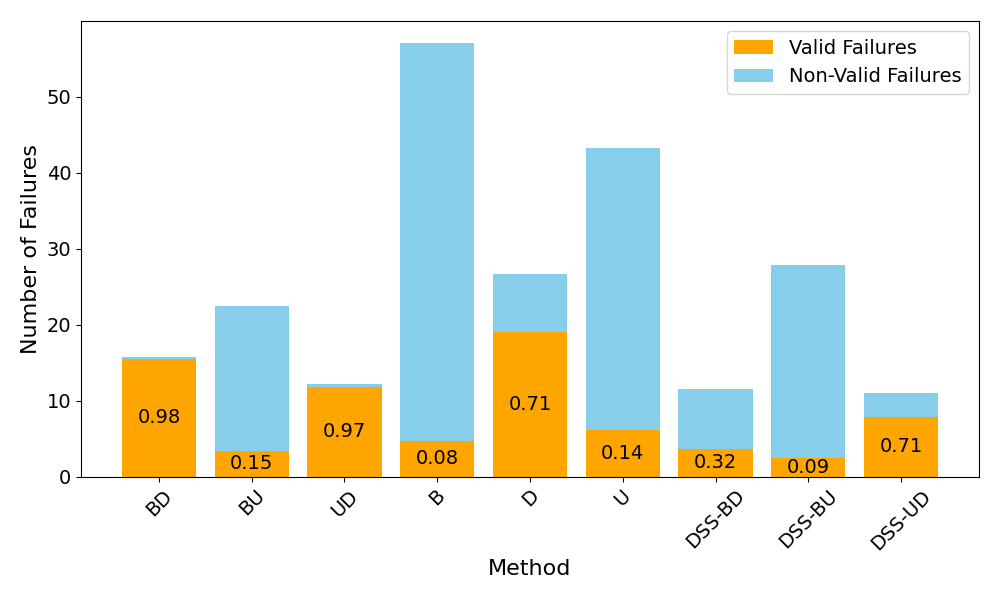}
        \caption{\vit}
    \end{subfigure}
    
    \vspace{0.5em}
    
    \begin{subfigure}{0.85\linewidth}
        \includegraphics[width=\linewidth]{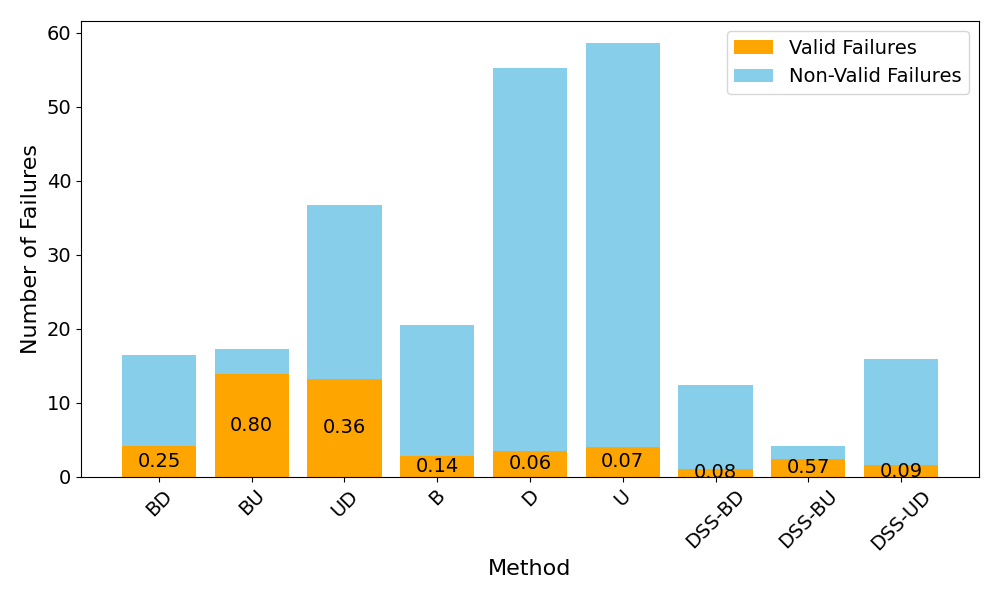}
        \caption{\ticp}
    \end{subfigure}
    
    \caption{Validity rate (\textit{valid\_rate}) and number of valid failures (\textit{n\_valid}) identified for \msim, \dss, and \ssim averaged for \davetwo, \vit, and \ticp. The average validity rate is shown in each bar plot.}
    \label{fig:rq1-stack-combined}
\end{figure}

The results for RQ\textsubscript{1} are shown in \autoref{fig:rq1-stack-combined}. The figure depicts the average number of failures and the number of \textit{valid} failures, identified from 10 runs for \msim, \ssim and \dss for the system \davetwo, \blue{and from 6 runs for the systems \vit and TCP}. In addition, in each bar plot the validity rate is annotated in the orange bar.

Across all configurations and system under tests, single-simulator (\ssim) approaches generally detect more total failures than multi-simulator ones, but many of these are invalid. The highest number of valid failures is achieved by \ssim on Udacity (U), which is comparable to the best multi-simulator configuration, BD for \davetwo.

While \dss yields in general fewer total failures overall, the number of valid failures remains for \davetwo similar to \msim. However, for \vit and \ticp the number of valid failures is in general lower for \dss.

Simulator-wise, for \davetwo \blue{and ViT}, BD produces fewer total failures than its single-simulator counterparts (B and D) but maintains a comparable number of valid ones. In contrast, BU identifies a similar total number of failures as B and U but significantly fewer valid cases, a pattern also observed for DSS-BU and UD.
\blue{For \ticp, the configurations BU and UD reach a significantly higher number of valid failures compared the single simulator-based counterparts, as well as to \dss and the multi-simulator configuration BD.}

The results in \autoref{fig:rq1-stack-combined} show that for \davetwo \blue{and ViT}, BD achieves the highest validity rate across all compared approaches and configurations, (99\% for \davetwo, \blue{98\% for ViT}). For \davetwo, BU has a validity rate of 55\%, which is surpassed by B, being close to that of DSS-BU. For UD the validity rate is higher compared to those of D, U, and DSS-UD.

\blue{For \ticp, BU has the highest validity rate (80\%), which is higher then the corresponding \dss based counterpart. Also, we can observe that all \ssim approaches achieve lower validity rates then the worst \msim combination BD (0.25).}

We further evaluated the convergence behavior of \msim and \ssim using the Hypervolume (HV) quality indicator, a widely adopted metric in search-based software engineering for assessing both the optimality and spread of Pareto fronts~\cite{LiEvaluateSBST22,LiQuality2019}. The HV analysis (available in our replication package~\cite{replication-package}) shows that \ssim configurations quickly reach stable HV values, whereas \msim achieves comparable values that continue to improve slightly over time. This suggests that extending the execution time of \ssim would likely not yield further gains in effectiveness, while \msim maintains gradual improvements as the search progresses.

\begin{table}[t]
\centering
\caption{Effectiveness: Statistical tests (Wilcoxon and Vargha-Delaney) for the comparison between \msim and \ssim, as well as between \msim and \dss for the metrics validity rate and number of valid failures. The letter L represents a large effect size magnitude, M represents a medium effect, and S represents a small effect (repeated comparisons are excluded (N/A)). Effect values are annotated with a star if the first approach of the comparison yields lower values than the compared approach.}
\label{tab:rq1-combined-stat}
\Huge
\resizebox{\textwidth}{!}{
\begin{tabular}{@{}lccccccccc@{}}
\toprule
& \multicolumn{1}{c}{\sc U} 
& \multicolumn{1}{c}{\sc B} 
& \multicolumn{1}{c}{\sc D} 
& \multicolumn{1}{c}{\sc BU}
& \multicolumn{1}{c}{\sc UD}
& \multicolumn{1}{c}{\sc DSS-BD} 
& \multicolumn{1}{c}{\sc DSS-BU} 
& \multicolumn{1}{c}{\sc DSS-UD} \\
& \( p \) (effect) & \( p \) (effect) & \( p \) (effect) & \( p \) (effect) & \( p \) (effect) & \( p \) (effect) & \( p \) (effect) & \( p \) (effect) \\
\midrule

\multicolumn{9}{l}{\textbf{DAVE-2}} \\

BD \\ 
\quad valid\_rate      & \textbf{$\sim$0} (L) & \textbf{$\sim$ 0} (L) & \textbf{$\sim$ 0} (L)  & \textbf{0.01} (L) & \textbf{$\sim$~0} (L) & \textbf{0.03} (L) & \textbf{0.03} (L) & 0.14 (-) \\
\quad n\_valid         & 1.00 (-) & 0.85 (-) & 0.83 (-)  & \textbf{0.03} (L) & 0.77 (-) & \textbf{0.03} (L) & \textbf{0.03} (L) & 0.06 (S) \\

BU \\ 
\quad valid\_rate      & 0.11 (-) & 0.77 (-) & 0.43 (-) & 0.77 (-) & \textbf{0.02} (L) & N/A &  0.56 (L*) & 0.77 (-) (-) \\
\quad n\_valid         & \textbf{0.04} (L) & \textbf{0.01} (L) & N/A &  0.16 (-)& 0.11 (-) & 1.00 (M) & 0.31 (M) & 0.08 (-) \\

UD \\ 
\quad valid\_rate      & \textbf{0.03} (L) & 0.92 (-) & 0.32 (-) & N/A & N/A & 0.16 (M) & \textbf{0.03} (L) & 0.14 (-) \\
\quad n\_valid         & 0.85 (-) & 0.77 (-) & 0.68 (-) & N/A & N/A & 0.16 (L) & 0.09 (M) & 0.44 (S) \\

\midrule

\multicolumn{9}{l}{\textbf{\blue{ViT}}} \\

BD \\ 
\quad valid\_rate      & \textbf{0.03} (L) & \textbf{0.03} (L) & \textbf{0.03} (L) & \textbf{0.03} (L) & 0.59 (S) & \textbf{0.04} (L) & \textbf{0.03} (L) & 0.14 (-) \\
\quad n\_valid         & 0.09 (M) & \textbf{0.03} (L) & 0.56 (S) & 0.06 (L) & 0.31 (M) & \textbf{0.03} (L) & \textbf{0.03} (L) & 0.06 (S) \\

UD \\ 
\quad valid\_rate      & \textbf{0.03} (L) & \textbf{0.03} (L) & \textbf{0.03} (L) & \textbf{0.03} (L) & N/A & 0.16 (M) & \textbf{0.03} (L) & 0.14 (-) \\
\quad n\_valid         & 0.22 (M) & 0.08 (M) & 0.31 (M) & 0.16 (M) & N/A & 0.16 (L) & 0.09 (M) & 0.44 (S) \\

BU \\ 
\quad valid\_rate      & 0.84 (-) & \textbf{0.03} (S) & \textbf{0.03} (L*) & N/A & N/A & 1.00 (M) & 0.31 (M) & 0.06 (-) \\
\quad n\_valid         & 0.44 (S) & 0.31 (M) & \textbf{0.04} (L*) & N/A & N/A & 1.00 (M) & 0.50 (-) & 0.08 (-) \\

\midrule

\multicolumn{9}{l}{\textbf{\blue{TCP}}} \\
BD \\
\quad valid\_rate & \textbf{0.04} (L) & 0.69 (M) & \textbf{0.04} (L) & 0.56 (S*) & 0.06 (L*) & \textbf{0.04} (L) & 0.14 (L) & 0.08 (L*) \\
\quad n\_valid    & 0.78 (L) & 0.59 (S) & 1.00 (-) & 0.09 (L*) & 0.06 (L*) & 0.07 (L*) & \textbf{0.04} (L) & 0.36 (M*) \\

UD \\
\quad valid\_rate & 0.06 (L*) & 0.16 (L*) & 0.06 (L*)  & 0.09 (L*) &  N/A& 0.09 (L) & \textbf{0.04} (L) & 0.56 (S) \\
\quad n\_valid    & 0.09 (L*) & 0.06 (L*) & 0.22 (L*)  & 1.00 (-) & N/A& 0.06 (L) & \textbf{0.04} (L) & 0.06 (L*) \\

BU \\
\quad valid\_rate & \textbf{0.04} (L) & 0.06 (L*) & \textbf{0.04} (L) & N/A & N/A & \textbf{0.04} (L) & \textbf{0.04} (L) & 0.29 (L*) \\
\quad n\_valid    & \textbf{0.04} (L) & 0.06 (L*) & \textbf{0.04} (L) & N/A & N/A & \textbf{0.04} (L) & \textbf{0.04} (L) & 0.06 (L*) \\
\bottomrule
\end{tabular}}
\end{table}

\textit{Statistical Test.} To assess whether the differences observed are statistical significant, we use the non-parametric pairwise Wilcoxon rank sum test (significance level 0.05) and the Vargha Delaney's $\hat{A}_{12}$ effect size to compare the \textit{valid\_rate} and the \textit{n\_valid} values shown in \autoref{fig:rq1-stack-combined}. We adopt the following standard classification for effect size values: an effect size $e$ is small, when 0.56 $\leq$ $e$ $<$ 0.64 or 0.36 $<$ $e$ $\leq$ 0.44, it is medium when 0.64 $\leq$ $e$ $<$ 0.71 or 0.29 $<$ $e$ $\leq$ 0.36 and large when $e$ $\geq$ 0.71 or $e$ $\leq$ 0.29. Otherwise, the effect size is negligible. The results are shown in \autoref{tab:rq1-combined-stat}.

For \davetwo, the statistical test results show that BD yields significantly higher validity rates with large effect sizes than \ssim, \msim and \dss approaches DSS-BU and DSS-UD. As for BU, the validity rate results for \msim are in general not better or worse than \ssim or \msim. For UD, the results are only significant compared to \ssim with Udacity (i.e., U).
Regarding the number of valid failures, we see that BD outperforms \dss-BD with a medium effect size, while having a similar validity rate. Regarding BU, \ssim B and U are significantly better than BU with a large effect size, what is consistent with the results in \autoref{fig:rq1-stack-combined}.
Concerning UD, we do not observe a statistical significant difference w.r.t. \ssim and \dss.


\blue{For \vit, similarly there is a statistically significant difference with large effect sizes for the valid rate between \msim-based and \ssim based approaches as well as the corresponding \dss-BD baseline. Also for UD there is a statistically relevant difference compared to \ssim approaches which is inline with the results with \autoref{fig:rq1-stack-combined}. For the number of valid failures, we see statistically significant higher values for BD compared to the single simulator approach B as well to corresponding DSS counterparts.}


\blue{For \ticp, both BD and BU configurations exhibit significantly higher validity rates and numbers of valid failures than the single-simulator (U, D) and corresponding \dss configurations, with large effect sizes. UD also shows statistically significant improvements over most \ssim configurations, though with smaller effects. Overall, results consistently confirm that combining complementary simulators enhances the discovery of valid failures across architectures, with BD emerging as the most effective configuration.}

\begin{tcolorbox}
$RQ_1$ (effectiveness). \blue{ Across all ADAS}, \msim achieves higher validity rates and numbers of valid failures than both \ssim and \dss (70\% vs.\ 65\% and 47\%). Among multi-simulator settings, BD (BeamNG–Donkey) performs best for DAVE-2 \blue{and ViT} (up to 99\% validity), \blue{while BU (BeamNG–Udacity) excels for TCP with up to 80\%}. This indicates that the most effective simulator pairing depends on the SUT: BD benefits vision- and control-based systems, whereas BU better captures perception-driven failures. Statistical tests confirm that \msim significantly outperforms both \ssim and \dss, except for \dss-BD, where it achieves comparable validity but a higher number of failures.
\end{tcolorbox}

\subsubsection{Efficiency (RQ\textsubscript{2})}

\autoref{tab:rq2-first-valid} shows the efficiency results, specifically the ratio of the search budget and its standard deviation when the first valid failure is detected.

For \davetwo we observe that all \msim configurations exhibit higher percentages for the first valid failure compared to \ssim, as expected. Compared to \dss, we see that the configuration BD achieves a better result than \dss-BD. However, across all configurations, the standard deviations for \msim are higher than those of \ssim or \dss. The combination UD is surpassed by \ssim and \dss approaches, beside DSS-BD which yields a higher search time. For BU we can observe a similar behavior.

The statistical test results are shown in \autoref{tab:rq2-stats}. While for \davetwo D significantly outperforms BD, BD significantly performs better then \dss-BD. Regarding BU, we can only observe that it is outperformed by \dss-BU and D, while there is no significant difference with the other approaches.
The \ssim configurations U and D yield significantly lower search times then UD, with large effect sizes. However, we do not observe a statistical significant difference of UD w.r.t. \dss-UD, which confirms our observations from \autoref{tab:rq2-stats}. 


\blue{For \vit, we can see that, similarly as for \davetwo, \msim combinations achieve on average higher values than \ssim-based search and \dss.}

\blue{For \ticp, \msim achieves lower scores with similar variation compared to \ssim. However the difference is not  statistically significant except for the comparison between BU and B for \ticp, where \msim achieves significantly better results.}

\begin{table}[t]
    \centering
    \caption{\changed{Efficiency results, i.e., time to first valid failure for \davetwo, \vit, and \ticp. Mean and standard deviation (std) are in per cent.}}
    \label{tab:rq2-first-valid}
    \begin{tabular}{lcccccc}
        \toprule
        \multirow{2}{*}{\textbf{Method}} & 
        \multicolumn{2}{c}{\textbf{\davetwo}} & 
        \multicolumn{2}{c}{\textbf{\blue{\vit}}} &
        \multicolumn{2}{c}{\textbf{\blue{\ticp}}} \\
        \cmidrule(lr){2-3} \cmidrule(lr){4-5} \cmidrule(lr){6-7}
         & {\small \textbf{Mean}} & {\small \textbf{Std}} 
         & {\small \textbf{Mean}} & {\small \textbf{Std}} 
         & {\small \textbf{Mean}} & {\small \textbf{Std}} \\
        \midrule
        BD     & 39.7 & 22.9 & 34.2 & 9.4 & 40.6   & 31.2   \\
        UD     & 35.3 & 24.0 & 56.5 & 22.2 & 31.0   & 33.3   \\
        BU     & 55.6 & 29.4 & 60.1 & 29.4 & 30.6   & 31.2   \\
        \midrule
        U      & 22.8 & 16.1 & 44.5 & 37.3 & 59.0   & 31.1   \\
        B      & 25.5 & 11.3 & 51.4 & 31.1 & 79.5   & 26.9   \\
        D      & 22.4 & 14.1 & 37.0 & 20.9 & 74.4 & 34.4   \\
        \midrule
        DSS-BD & 56.6 & 16.3 & 37.1 & 2.2 & 61.1  & 12.5   \\
        DSS-UD & 39.9 & 13.6 & 43.4 & 6.5 & 45.8   & 63.5   \\
        DSS-BU & 33.5 & 13.2 & 34.6 & 5.2 & 51.6   & 10.2   \\
        \bottomrule
    \end{tabular}
\end{table}

\changed{
\begin{figure}[h!]
    \centering

    \begin{subfigure}[b]{0.328\linewidth}
        \centering
        \includegraphics[width=\linewidth]{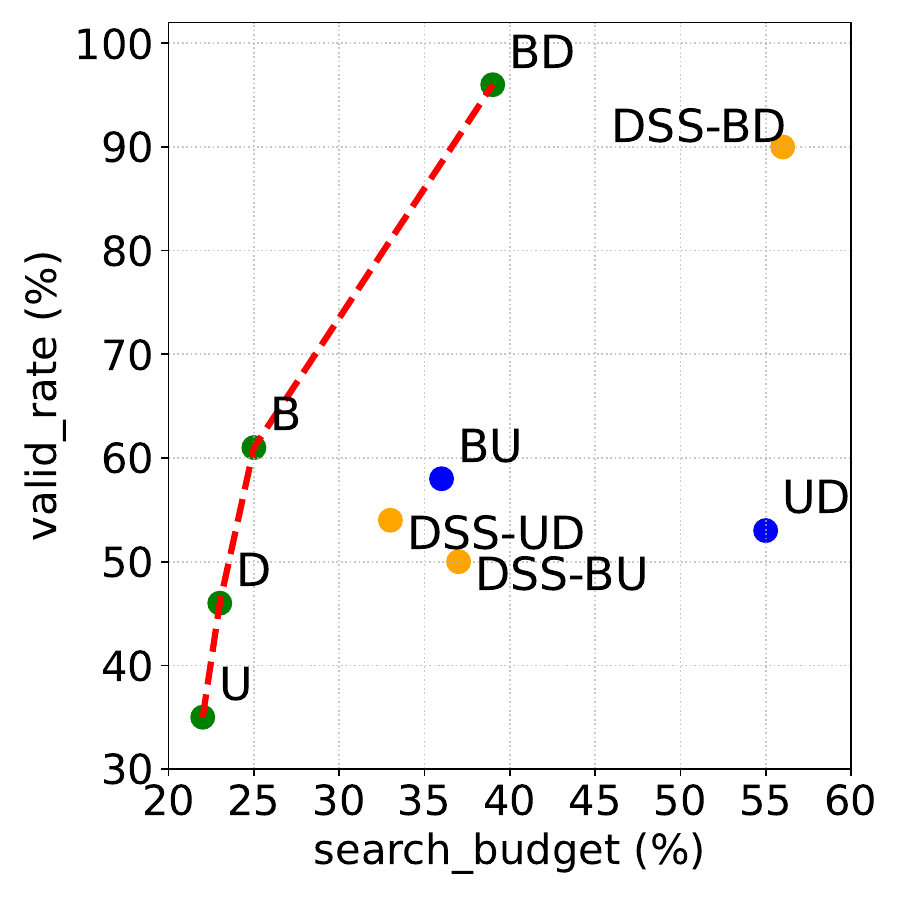}
        \caption{\davetwo}
        \label{fig:pareto-cs1}
    \end{subfigure}
    \hfill
    \begin{subfigure}[b]{0.328\linewidth}
        \centering
        \includegraphics[width=\linewidth]{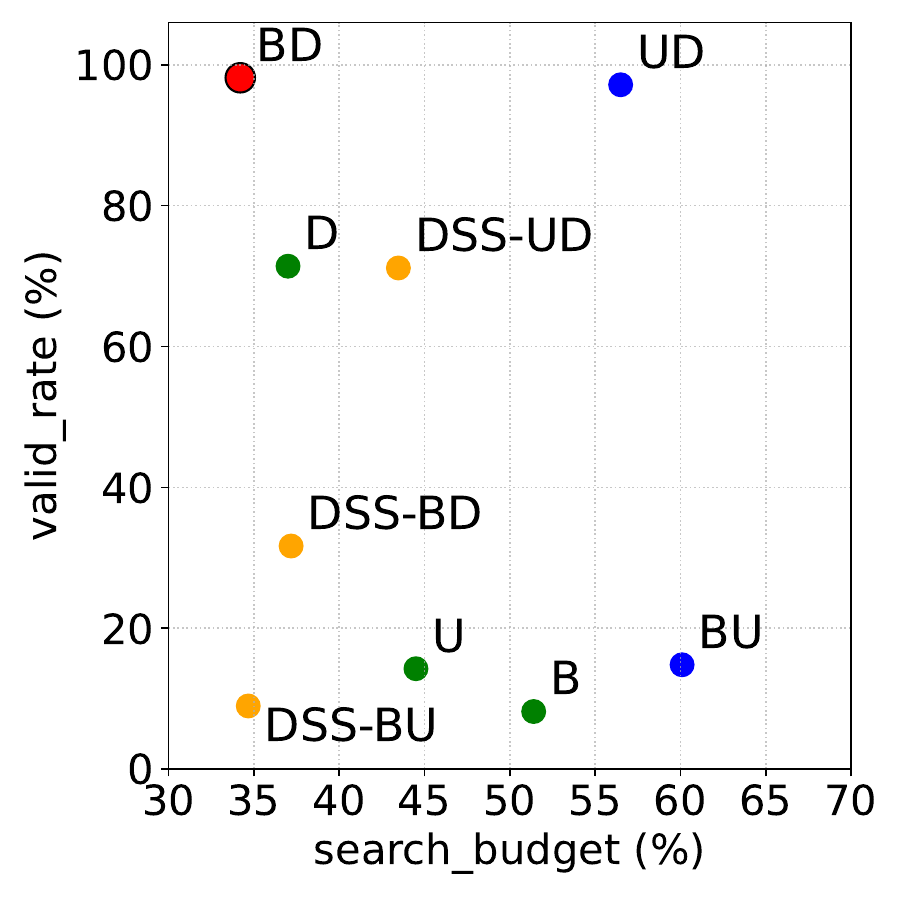}
        \caption{\vit}
        \label{fig:pareto-cs2}
    \end{subfigure}
    \hfill
    \begin{subfigure}[b]{0.328\linewidth}
        \centering
           \includegraphics[width=\linewidth]{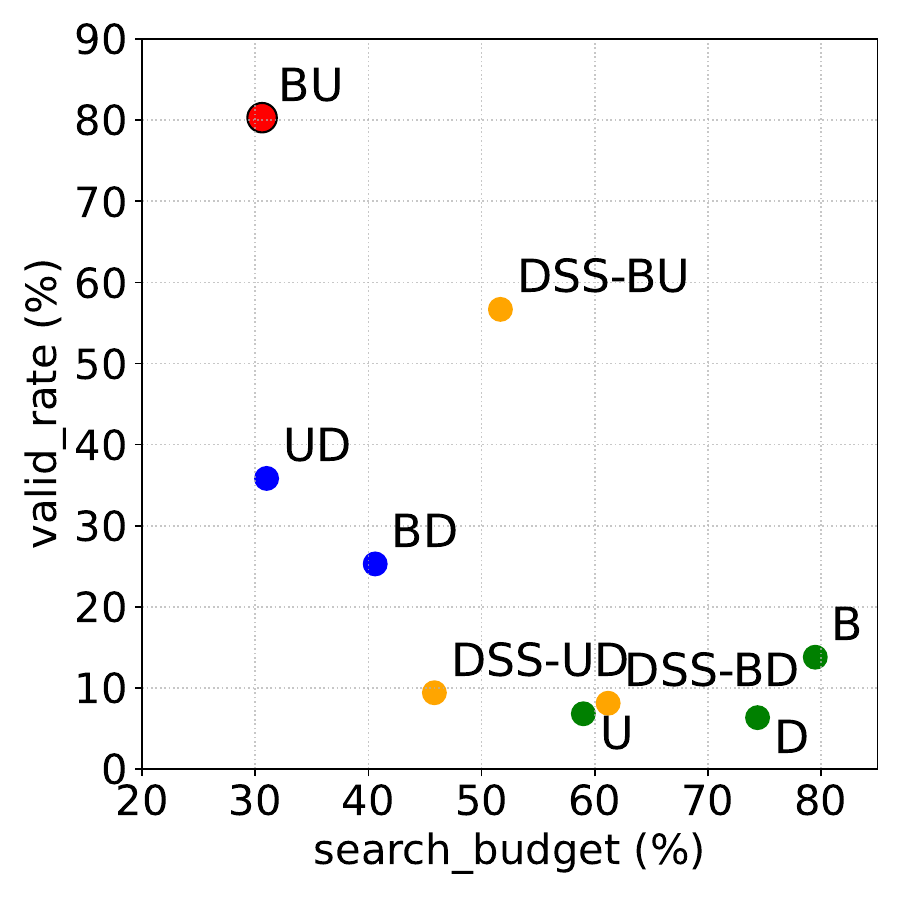}
        \caption{\ticp}
        \label{fig:pareto-cs3}
    \end{subfigure}

    \caption{Visualization of the tradeoff between efficiency (first valid failure) and effectiveness (\textit{valid\_rate}) for \msim, \dss and \ssim for different configurations across three case studies. The Pareto front/point is visualized in red.}
    \label{fig:rq2-pareto-all}
\end{figure}
 }
\begin{table}[t]
\centering
\caption{\changed{(\davetwo) Efficiency. Statistical tests (Wilcoxon and Vargha-Delaney) for identifying the first valid failure (repeated comparisons are excluded (N/A)) for the ADAS \davetwo, \vit and \ticp. Effect size magnitudes are annotated with a star, if the first approach (rows) of the comparison yields lower values than the second approach (columns).}}
\label{tab:rq2-stats}
\Huge
\resizebox{\textwidth}{!}{
\begin{tabular}{@{}lllllllll@{}}
\toprule
& \multicolumn{1}{c}{\sc U} 
& \multicolumn{1}{c}{\sc B} 
& \multicolumn{1}{c}{\sc D} 
& \multicolumn{1}{c}{\sc UD}
& \multicolumn{1}{c}{\sc BU}
& \multicolumn{1}{c}{\sc DSS-BD} 
& \multicolumn{1}{c}{\sc DSS-BU} 
& \multicolumn{1}{c}{\sc DSS-UD} \\
& \( p \) (effect) & \( p \) (effect) & \( p \) (effect) & \( p \) (effect) & \( p \) (effect) &  \( p \) (effect) & \( p \) (effect) & \( p \) (effect)   \\
\midrule
\multicolumn{9}{l}{\textbf{\davetwo}} \\

BD \\ 
\quad     & 0.11 (-) & 0.06 (-) & \textbf{0.03} (L*) & 0.70 (-) & 0.16 (-) & 0.01 (L*) & 0.38 (-) & 1.00 (-) \\

BU \\ 
\quad   & 0.16 (-) & 0.28 (-) & \textbf{0.04} (M*) & 0.19 (-) & N/A & 0.85 (-) & \textbf{0.03} (L) & 0.16 (-) \\

UD \\ 
\quad   & \textbf{0.03} (L*) & \textbf{0.02} (L*) & \textbf{0.01} (L*)  & N/A & N/A &  0.01 (L*)  & 0.92 (-) & 0.49 (-) \\

\midrule
\multicolumn{9}{l}{\textbf{\blue{\vit}}} \\

BD \\
\quad     & 1.00 (L) & 0.31 (S) & 0.84 (S) & 0.16 (L*) &  0.16 (L) & 0.06 (L) & 0.84 (-) & 0.09 (L) \\

BU \\
\quad     & 0.56 (M) & 0.56 (S) & 0.44 (L) & N/A & 0.69 (-) &  0.09 (L) &   0.16 (L) & 0.31 (M) \\

UD \\
\quad     & 0.56 (M) & 0.84 (S) & 0.31 (L) & N/A & N/A & 0.16 (L) &  0.16 (L) &  0.44 (S) \\

\bottomrule
\multicolumn{9}{l}{\textbf{\blue{\ticp}}} \\

BD \\
\quad & 0.22 (M*) & 0.16 (L*) & 0.09 (L*) & 0.31 (L) & 0.44 (L) & 0.31 (L*) & 0.56 (L*) & 0.56 (L*) \\

BU \\
\quad  & 0.16 (L*) & \textbf{0.04} (L*) & 0.16 (L*) & N/A & N/A & 0.09 (L*) &  0.31 (L*) & 0.44 (L*)\\

UD \\
\quad & 0.22 (L*) & 0.16 (L*) & 0.16 (L*) & N/A & 1.00 (-) & 0.06 (L*) & 0.44 (L*) & 0.44 (L*) \\

\bottomrule
\end{tabular}}
\end{table}

\changed{
\autoref{fig:rq2-pareto-all} shows the study results considering both effectiveness and efficiency (validity rate vs. search budget required for first valid failure) for \davetwo, \vit and \ticp for \msim, \ssim and \dss. Particuarly, for \davetwo the methods whose values are non-dominated either in effectiveness nor efficiency are BD, B and D (Pareto-optimal solutions). \ssim, when using BeamNG and Donkey, is more efficient in finding the first valid failure, while based on the results of the \msim combination BD, \msim is significantly more effective.
For \vit/\ticp, we see that the \msim combination BD/BU is dominating all remaining configurations yielding both a high validity rate with a low search effort for the first valid failure.
}
\begin{tcolorbox}
$RQ_2$ (efficiency). Across all configurations, \msim achieves efficiency comparable to \ssim, requiring a similar search budget to find the first valid failure. Among multi-simulator settings, BD and UD are the most efficient for \davetwo \blue{and \textit{ViT}, while BU performs best for TCP}. Compared to \dss, in general \msim shows better efficiency, whereas \dss-BU remains in two out of three comparisons slightly more time-efficient than BU. Overall, \msim maintains competitive efficiency while yielding substantially higher failure validity.
\end{tcolorbox}

\subsubsection{Prediction (RQ\textsubscript{3})}
\label{sec:rq3}
\changed{To reduce \textit{unnecessary} simulator evaluations, we extend \msim with a \emph{surrogate disagreement predictor}.
This surrogate is a classifier, trained using \textbf{labeled outcomes} of previous runs, where each instance is annotated as an \emph{agreement} or \emph{disagreement}. The surrogate outputs the probability of a \emph{disagreement event} between simulator pairs.
Specifically, we gathered all test cases from five completed runs using BD, along with five additional runs using a modified version of \msim focused on identifying disagreements (\autoref{sec:metrics}). In total, we collected a dataset comprising 580 agreements and 572 disagreements for \davetwo, 581 agreements and 581 disagreements for \vit, and 383 agreements and 383 disagreements for TCP. These datasets were then used to train five classification models aimed at predicting disagreement occurrences. We train and validate multiple surrogate classifiers (Decision Tree, Random Forest, SVM, Logistic Regression, and Gradient Boosting) using 5-fold cross-validation with an 80:20 split. As reported in \autoref{tab:rq3-cv-results-updated}, the Random Forest model achieved the best performance for every system under test, with F-1 scores close to or above 80\%.
We therefore integrate this model into \msim as the surrogate predictor.}

\changed{During the test generation phase, each newly generated candidate scenario is first evaluated by the surrogate model.
If the predicted probability of simulator disagreement exceeds a threshold ($\tau=0.7$), the candidate is discarded before running the full simulation.
Otherwise, it is executed across both simulators.
This filtering mechanism allows \msim to save simulation budget by focusing exploration on high-value, disagreement-prone regions of the search space.}

\changed{Across all three case studies, integrating the surrogate disagreement predictor into \msim (\textbf{BD-P}) improved the overall test generation effectiveness. 
The surrogate-guided runs produced a higher number of valid and fault-revealing test cases compared to the baseline configuration (BD), increasing the mean \textit{n\_valid} and \textit{valid\_rate} by approximately 38\% on average. 
This confirms that the surrogate model not only reduces redundant simulator queries but also helps focusing the search on higher-yield regions of the input space, maintaining an average validity rate above 90\% across runs.}

\changed{The use of the surrogate model also leads to efficiency improvements in two out of the three case studies. For \davetwo it yields an improvement of 14\%, and 2\% for \ticp. These results indicate that surrogate-based filtering can substantially accelerate simulation-based testing while maintaining the ability to identify a broad range of failure cases.}

\begin{table}[t]
\centering
\caption{\blue{Classifier results after 5-fold cross-validation (in \%) for predicting disagreements in three simulator configurations. The best results are marked in bold.}}
\label{tab:rq3-cv-results-updated}
\resizebox{\columnwidth}{!}{
\begin{tabular}{lcccccc}
\toprule
\multirow{2}{*}{\textbf{}} & 
\multicolumn{2}{c}{\textbf{\davetwo}} & 
\multicolumn{2}{c}{\textbf{\blue{\vit}}} & 
\multicolumn{2}{c}{\textbf{\blue{TCP}}} \\
\cmidrule(lr){2-3} \cmidrule(lr){4-5} \cmidrule(lr){6-7}
 & \textbf{F-1} & \textbf{AUC-ROC} 
 & \textbf{F-1} & \textbf{AUC-ROC}
 & \textbf{F-1} & \textbf{AUC-ROC} \\
\midrule
DT      & 79 & 78 & 58 & 65 & 60 & 65 \\
RF      & \textbf{83} & \textbf{89} & \textbf{74} & \textbf{81} & \textbf{78} & \textbf{86} \\
SVM     & 69 & 71 & 65 & 77 & 58 & 80 \\
LREG    & 57 & 62 & 58 & 65 & 60 & 65 \\
GBOOST  & 80 & 87 & 73 & 80 & 75 & 82 \\
\bottomrule
\end{tabular}
}
\end{table}

\begin{table}[t]
\centering
\caption{\changed{Mean and standard deviation of effectiveness and efficiency metrics for BD-P and BD approaches across three use cases, averaged over 10 runs for \davetwo, 6 runs for \vit, and 6 runs for \ticp.}}
\label{tab:rq3-summary}
\resizebox{\columnwidth}{!}{
\begin{tabular}{lcccccc}
\toprule
\multirow{2}{*}{\textbf{Metric}} 
& \multicolumn{2}{c}{\textbf{\davetwo}} 
& \multicolumn{2}{c}{\textbf{\blue{\vit}}} 
& \multicolumn{2}{c}{\textbf{\blue{\ticp}}} \\
\cmidrule(lr){2-3} \cmidrule(lr){4-5} \cmidrule(lr){6-7}
 & \textbf{BD-P} & BD & \textbf{BD-P} & BD & \textbf{BD-P} & BD \\
\midrule
n\_valid & 25.3 $\pm$ 7.3 & 23.0 $\pm$ 13.1 & 15.8 $\pm$ 2.0 & 15.5 $\pm$ 8.3 & 8.3 $\pm$ 3.6 & 4.1 $\pm$ 4.8 \\
valid\_rate (\%) & 92.9 $\pm$ 7.9 & 98.5 $\pm$ 3.0 & 97.9 $\pm$ 3.0 & 98.1 $\pm$ 4.1 & 44.3 $\pm$ 8.3 & 25.9 $\pm$ 18.7 \\
first\_fail (\%) & 34.0 $\pm$ 14.1 & 39.7 $\pm$ 22.9 & 43. $\pm$ 16.7 & 34.2 $\pm$ 9.4 & 39.8 $\pm$ 25.9 & 40.6 $\pm$ 31.2 \\
\bottomrule
\end{tabular}
}
\end{table}

\begin{tcolorbox}
\textit{RQ\textsubscript{3} (prediction).} 
Among all trained classifiers, the Random Forest achieved the best disagreement-prediction performance (AUC: 0.81–0.89, F-1: 0.74–0.83). Integrating this surrogate into \msim enabled the framework to bypass low-value simulations, improving both effectiveness and efficiency by increasing the number of valid failures and reducing variability compared to baseline approaches.
\end{tcolorbox}

\subsection{Qualitative Analysis}\label{sec:qualitative}

In the following, we describe a qualitative analysis concerning our evaluation results. In the first part, we highlight disagreement scenarios encountered with \msim. In the second part, we report on the observations we made regarding results using the disagreement classifier from RQ\textsubscript{3}.

\subsubsection{Disagreements} We manually analyzed the  disagreements found by \msim across different configurations. We made following observations.

\begin{figure}[t]
\centering

\begin{subfigure}[b]{0.48\textwidth}
    \centering
    \includegraphics[height=4.5cm]{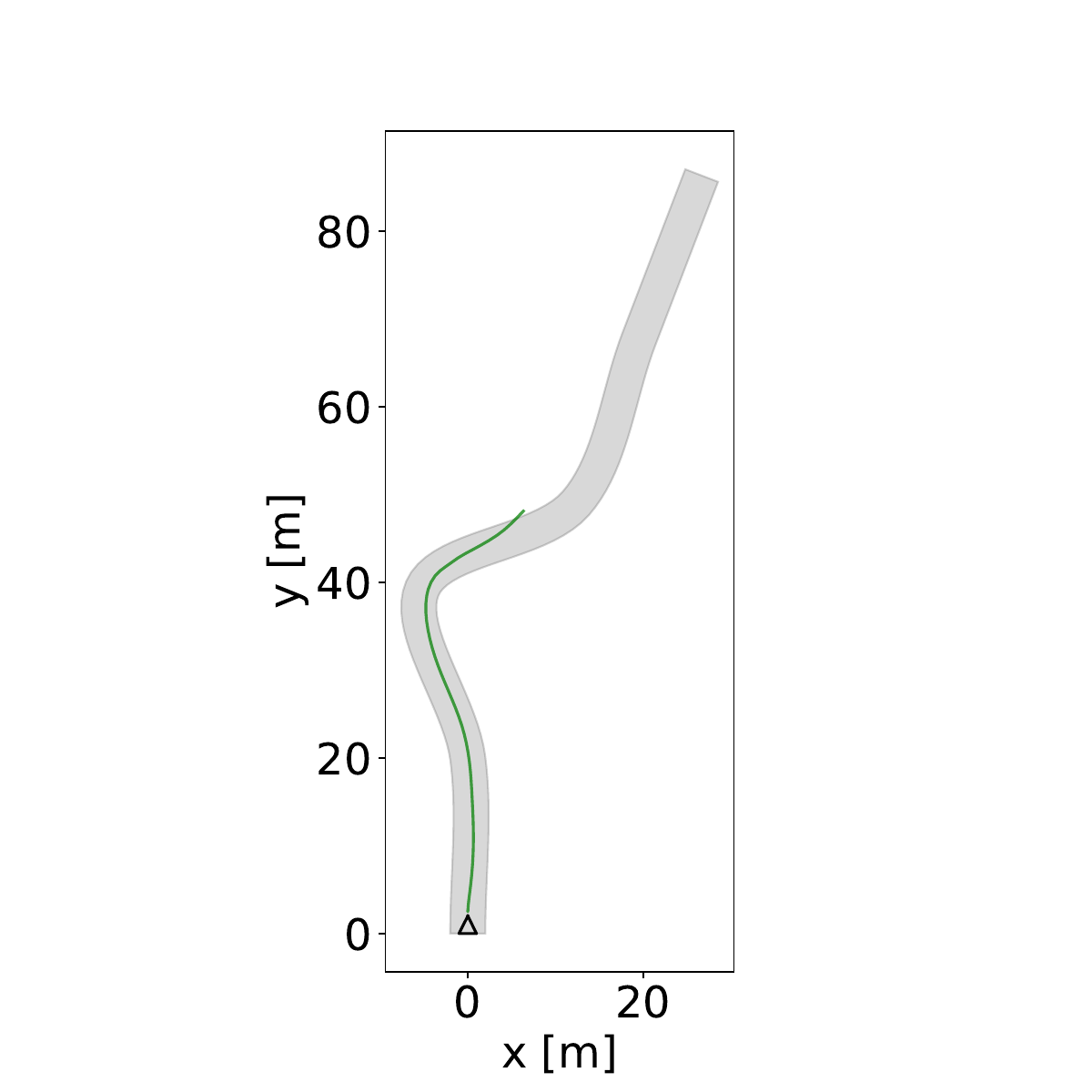} 
    \label{fig:disagree-short-bng}
\end{subfigure}
\begin{subfigure}[b]{0.48\textwidth}
    \centering
    \includegraphics[height=4.5cm]{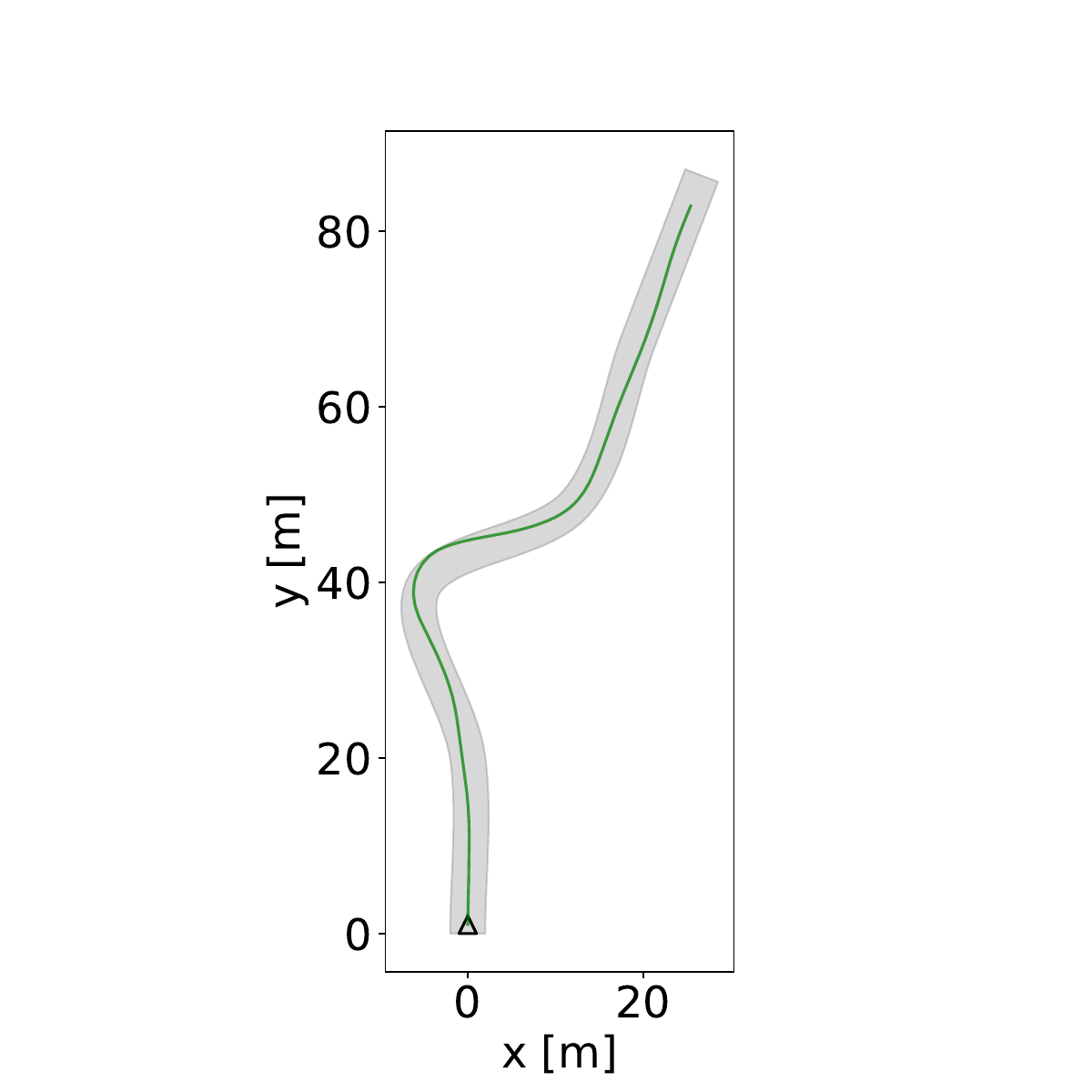} 
    \label{fig:disagree-short-dnk}
\end{subfigure}

\caption{Example of a disagreement scenario. Left: Vehicle simulated in BeamNG leaves short segment heading towards following segment. Right: Vehicle simulated in Donkey stays within the lane.}
\label{fig:qualitative-obs1}

\end{figure}

In the BeamNG simulator for \davetwo occasionally, the vehicle switches to the left lane whenever the underlying segment is short, i.e., $<$ 15m, and the following segment is a left turn (\autoref{fig:qualitative-obs1}). The car continues driving on the left lane towards the next segment, incurring in a failure. 

\changed{A similar behavior was observed for \ticp in Donkey and Udacity for sharp and short left turns followed by a sharp right turn, where the vehicle did not continue driving to the left, but proceeded straight to align with the subsequent path after the right turn. This behaviour could be explained by the fact that the vehicle observes the upcoming segment, and considers this shortcut as the immediate straight continuation of its path. However, in the remaining simulators and system under test configurations the vehicle stays in the lane, even when driving on short segments.

Specifically to \ticp, we observed that the model exhibits partially in all simulations environments a slightly higher XTE variation compared to \davetwo and \vit even when the single-simulator based failure count is less then for \davetwo or \vit. At the same time we could see that the model steers in general more extreme in curves compared to \davetwo or \vit which could explain its robust performance and the lower failure rate compared to the other system under tests.}

Across all system under tests, we observed also that in the BeamNG simulator, the vehicle switches often to the left lane in the last but one segment after meeting a sharp curve. This behaviour might be related to the fact that the road terminates after the last segment. However, we did not observe a similar behaviour for the vehicle in the Donkey or Udacity simulators, which might be attributed to the visualization/rendering differences in the simulators.

\begin{figure}[h!]
    \centering
    \includegraphics[width=0.55\linewidth]{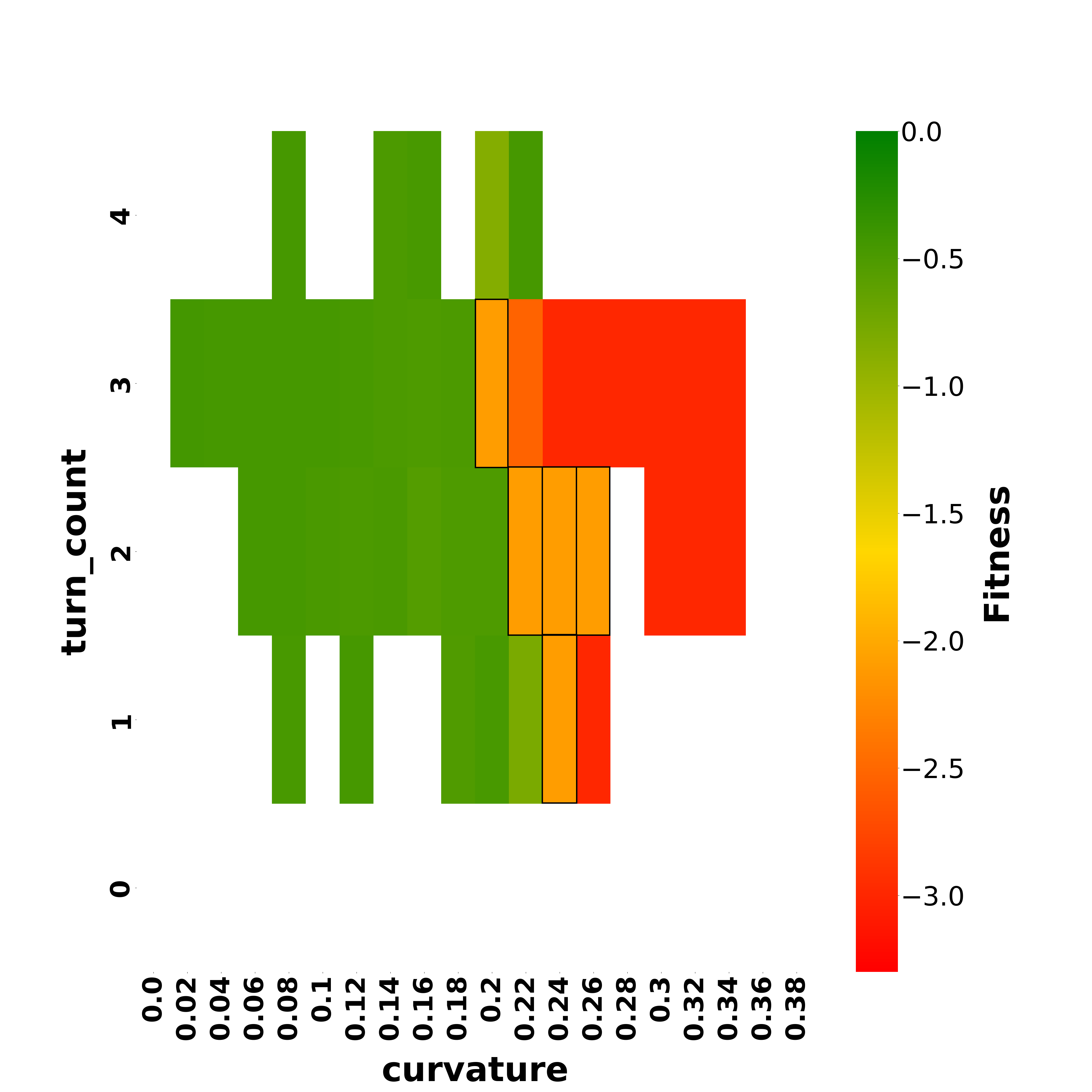}
    \caption{Feature map of tests found in one run with \msim employing disagreement classifier (BD-P) for \davetwo. Cells containing roads for which disagreements are predicted are marked in orange with a black box.}
    \label{fig:example-fmap-predict}
\end{figure}

\subsubsection{Feature Maps of Prediction-based Search}

When evaluating the results of \msim augmented with the prediction classifier, we inspected the feature maps before failure validation. We could observe that disagreements are predicted in regions which are between failing and non-failing cells of the map. An example for DAVE-2 is shown in \autoref{fig:example-fmap-predict}. Cells that contain tests that are predicted as disagreements by the classifier are marked as orange. This observation highlights the fact that our classifier is likely to perform accurately in predicting disagreements, as it is predicting disagreement on test cases which lie on the boundary between scenarios/roads that pass and fail. Similar situations were observed in the other simulator ensembles.

\section{Discussion}
\label{sec:discussion}

In this section, we discuss implications, insights and limitations related to our results.

\subsection{Approach} While our study utilizes two simulators in the ensemble, \msim can be extended to incorporate a larger number of simulators. To reach consensus on the criticality, we suggest to use an odd number of simulators. Further, the feasibility of the approach with multiple simulators is basically constrained by the underlying computational resources. To mitigate this, test executions in \msim with multiple simulators might be parallelized, independent of the underlying test algorithm used. 

\subsection{Lessons Learned} Our case study shows that employing multiple simulators in the search does improve the effectiveness of finding simulator-agnostic failures compared to single-simulator testing. 
However, in the literature, test case generation approaches in general report the evaluation results which are based on single-simulator executions. We suggest that failures found \textit{during} the test generation or at least the final identified failures, should be evaluated in a second simulator to assess whether the failures are likely to be simulator-agnostic.

\subsection{Root Cause Analysis} Our paper proposes a testing technique to identify failing test cases where simulators agree on. It is out of the scope of this study to investigate the underlying root causes of disagreements between simulator executions, as it would likely require to inspect, debug, and profile the simulator's code (in \autoref{sec:qualitative} we qualitatively analyzed two disagreement scenarios). Additionally, as outlined in RQ\textsubscript{3}, the \msim approach can be configured to search for tests on which simulators disagree to support the identification of the root causes behind failures. As shown by Jodat et al.~\cite{JodatSpurious23} and Ben Abdessalem et al.~\cite{Abdessalem-ICSE18}, decision trees or decision rules can be then applied on underlying roads and simulation traces to derive conditions on input variables and analyze the root causes behind inconclusive and simulator-specific failures.

Among the possible explanations for the disagreements, we conjecture that variations in vehicle dynamics may have influenced our results. Specifically, we observed that the vehicle in Udacity exhibited reduced steering capability compared to those in Donkey and BeamNG. This could account for the high number of simulator-specific failures in Udacity and explain why Udacity appeared more reliable for independent validation (in BD) than during the multi-simulator search, where it introduced spurious failures. As actionable feedback, we recommend incorporating or explicitly detailing vehicle characteristics~\cite{pan2023safetyassessmentvehiclecharacteristics} in scenario definitions for future scenario-based ADAS testing.

\changed{We observed that BD performs best for DAVE-2 and ViT, whereas BU excels for TCP. This can be explained by the differing sensitivities of the ADAS architectures to physical versus perceptual variability. BD combines simulators with heterogeneous vehicle dynamics (soft- vs. rigid-body physics), which better exposes control-related faults, while BU combines simulators with distinct rendering and camera models, thus better revealing perception-related inconsistencies. Consequently, it appears that the optimal simulator ensemble depends on the dominant failure modes of the SUT.}

\subsection{Validation} In our study, the validation was dependent on three hyperparameters: the number of re-executions, the failure rate threshold, and the number of failures selected per cell. We set the failure rate threshold to 100\%. However, in practice, the threshold for the validation can be configured based on the number of simulators used. We expect that the more diverse simulation environments are employed, the more difficult would be to achieve failure rates of 100\%. Regarding the number of re-executions we suggest to perform preliminary experiments with a different number of re-executions and compare the results, as conducted in our study. 
To select the number of selected failing tests per cell, we carried out preliminary experiments with a higher number of failures, i.e., 5. However, we did not observe significantly different evaluation results. Moreover, we did not select more tests per cell because the feature map already discretizes the search domain, making the diversity of roads within a single cell being likely low.

\changed{\subsection{Generalization of Our Approach}}

\changed{While our empirical evaluation focuses on lane-keeping assistance systems, the \msim framework is not limited to this specific ADAS function. The proposed architecture and workflow, based on simulator coupling, multi-objective search, and cross-simulator agreement analysis, can be generalized to other ADAS and ADS functionalities.}

\changed{First, the representation of tests is model-based and modular: road geometry is encoded through control points and spline interpolation, but this abstraction can be extended to other types of driving scenarios. For instance, intersections, pedestrian crossings, or obstacle avoidance can be represented using higher-level scene graphs or semantic maps instead of simple lane curves. Similarly, perception-heavy ADAS such as automatic emergency braking (AEB), adaptive cruise control (ACC), or lane-change assist could be evaluated by defining task-specific fitness functions (e.g., time-to-collision, minimum distance to obstacles, or stability of steering trajectories).}

\changed{Second, the optimization procedure is independent of the specific fitness formulation. Our search and repopulation strategies can operate over any domain where simulation feedback is measurable, making it possible to reuse the same search backbone for tasks with different control policies or sensing modalities (e.g., radar-, LiDAR-, or camera-based). This in particular is enabled by the modular architecture of the extended framework OpenSBT as shown in various case studies~\cite{sorokin2023opensbt}.}

\changed{Finally, the \msim framework could be extended to support system-level ADS validation by incorporating full autonomy stacks like Autoware~\cite{autoware} or Apollo~\cite{apollo}. In this context, cross-simulator disagreement analysis could help identify environment- or simulator-induced discrepancies in perception, planning, and control, offering a systematic means of analyzing simulation fidelity and testing robustness of the end-to-end driving system. However, since no existing autonomous driving stack is truly simulator-agnostic, extending \msim to full ADS validation remains an open engineering challenge that requires standardized interfaces and consistent scene representations across simulators.}

\section{Threats to Validity}\label{sec:threats}

\textit{External validity} concerns the extent to which our results can be generalized beyond the studied settings. We evaluated \msim on a well-established case study involving three lane-keeping models of increasing complexity, each tested across different simulator pairs drawn from a diverse set of environments, BeamNG, DonkeyCar, and Udacity.
\blue{It would be interesting to investigate whether the results generalize to other ADAS functions, such as automated emergency braking, or to higher levels of autonomy using full driving stacks like Autoware~\cite{autoware} or Apollo~\cite{apollo}. This would require testing in more complex urban simulation environments, such as AWSIM~\cite{awsim} or CARLA~\cite{carla}.
The main difficulty lies in the technical integration of modern autonomous driving stacks with multiple simulators, as these frameworks are typically coupled with a single specific simulator. Achieving cross-simulator consistency in terms of scenarios, visual appearance, simulation conditions, and interactions with other actors poses significantly greater challenges than in our current study.}

\textit{Internal validity} risks refer to confounding factors that could affect the interpretation of results. To mitigate these, we adopted several strategies. First, to avoid biased evaluations due to unaligned deployment conditions (e.g., communication delays between the SUT and the simulator), we conducted a preliminary validation of XTE variation across manually defined driving scenarios. Second, we aligned the simulated roads across all simulators, building upon existing research~\cite{matteoMaxibon2024}. Third, we defined validity constraints for generated roads and regenerated invalid samples after mutation.
For the BU setup, simulators were executed on different hardware configurations due to compatibility constraints (BeamNG requires running on a Windows operating system). This may have influenced BU results, although BU did not outperform other configurations in our experiments.
To ensure that detected failures were simulator-agnostic, we validated all failures identified by \msim, \dss, and \ssim on one or two simulators left out during the search. We also re-ran failing tests multiple times to filter out failures caused by transient or flaky simulation conditions.

\textit{Construct validity} threats relate to the adequacy of the chosen metrics. To evaluate performance, we used two primary metrics: the number of simulator-agnostic failures and the validity rate, i.e., the ratio between simulator-agnostic failures and validated failures. The first metric is a standard measure in test case generation studies~\cite{Kluck19Nsga2ADAS}, while the second captures the proportion of failures that remain consistent across simulators, reflecting robustness and transferability. Additionally, we analyzed the convergence of the Hypervolume (HV) indicator for \msim and \ssim, showing that \ssim converges earlier, while \msim continues improving slightly over time. We excluded \dss from HV analysis as it is not Pareto-based.

To foster \textit{replicability}, we publicly release the implementation of \msim, along with all experimental data for \msim, \ssim, and \dss. The search algorithms and result analyses are implemented using the modular and open-source framework OpenSBT~\cite{sorokin2023opensbt}, which supports easy integration of additional case studies for future research.

\section{Related Work}
\label{sec:related}

In this section, we outline two categories of related work. The first category targets ADAS testing approaches that use one simulator for test generation. The second category reports studies using multiple simulators for testing.

\subsection{Single-Simulator Approaches for ADAS Testing}

Most current test generation methods use search-based approaches to automatically create test cases for DNN-driven ADAS systems~\cite{moghadam2023machine,2020-Riccio-FSE,Abdessalem-ASE18-1,Abdessalem-ASE18-2,Abdessalem-ICSE18,deepxplore,deeptest,deeproad,Kim:2019:GDL:3339505.3339634}.  
In this domain, test cases consist of individual driving images or road topologies, which are rendered through a driving simulator. Abdessalem et al.~\cite{Abdessalem-ASE18-1,Abdessalem-ASE18-2,Abdessalem-ICSE18} integrate genetic algorithms with machine learning techniques to test a pedestrian detection system. Mullins et al.~\cite{MULLINS2018197} apply Gaussian processes to guide search-based test generation toward unexplored regions within the input space. Gambi et al.~\cite{asfault} leverage procedural content generation to propose a search-based test generation approach for ADAS.
Riccio and Tonella~\cite{2020-Riccio-FSE} introduce, a model-based test generator that leverages Catmull-Rom splines, the test representation as used in our approach, to define road shapes, producing test cases at the behavioral frontier of self-driving car models. 
Arrieta et al.~\cite{7969377} apply a genetic algorithm to generate tests for cyber-physical systems, optimizing them across three dimensions: requirement coverage, test case similarity, and execution time. 
Lastly, Lu et al.~\cite{9712397} employ reinforcement learning to discover environmental configurations that induce crashes. DeepQTest~\cite{lu2023deepqtesttestingautonomousdriving} is a testing approach that uses reinforcement learning to learn environment configurations with a high chance of revealing ADAS misbehaviors. In another work~\cite{epitester}, epigenetics algorithms are used to test ADAS in dynamically changing environments.

Among the fuzzing domain, DriveFuzz~\cite{drivefuzz} leverages the physical state of the vehicle, along with oracles grounded in real-world traffic rules, to guide the fuzzer toward uncovering potential misbehaviors. AutoFuzz~\cite{zhongETAL2021}, on the other hand, focuses on fuzzing the test scenario specifications. Before initiating the fuzzing process, it employs a seed selection mechanism using a binary classifier that identifies seeds with a higher likelihood of violating traffic rules. AV-Fuzzer~\cite{liETAL2020} applies a genetic algorithm, informed by the positioning of globally monitored non-player characters (NPCs) within each driving scenario. NPCs deemed to have a higher likelihood of safety violations~\cite{Jha2019MLBasedFI} are prioritized for evolution. Cheng et al.~\cite{10.1145/3597926.3598072} introduce BehaviorMiner, an unsupervised model that extracts temporal features from predefined scenarios and employs clustering-based abstraction to group behaviors with similar features into abstract states. 

All these approaches use a single-simulator approach, with some works validating their propositions on multiple simulators~\cite{10.1145/3597926.3598072}, for example, because different ADS are compatible/integrated only with specific simulation platforms~\cite{pan2023safetyassessmentvehiclecharacteristics}.
Our approach differs from these solutions because it uses an ensemble of simulations during search-based testing to retrieve more accurate fitness signals, based on the consensus among the simulators.

\subsection{Multi-Simulator Approaches for ADAS Testing}

A study by Borg et al.~\cite{Borg21CrossSimTesting} investigates the comparability of multiple simulators for testing a pedestrian vision detection system. The study evaluates a large set of test scenarios on both PreScan~\cite{prescan} and Pro-SiVIC~\cite{pro-sivic}. The study reports inconsistent results in terms of safety violations and behaviors across these simulators. Consequently, the authors suggest that a single-simulator approach for ADAS testing might be unreliable, especially when failures are highly dependent on the chosen simulator. Moreover, a recent study by Amini et al.~\cite{AminiFlaky2024} has analyzed the degree of flakiness affecting ADAS testing. The study evaluates several simulators and ADAS showing that test flakiness is common and can significantly impact the test results. The authors propose the usage of machine learning classifiers to identify flaky ADAS tests.
Another study by Wagner et al.~\cite{8814268} evaluates the translation of real-world driving scenarios to executed scenarios in a simulator. Their results show that 
a reprocessing error exists, which can be basically attributed to sensor model offsets and can be tackled by employing scenario-based sensor models.

To address simulator disagreements, Biagiola et al.~\cite{matteoMaxibon2024} involve search-based testing across multiple simulators, provided that the same test scenario and ADAS under test can be consistently represented. Their method combines the predicted failure probabilities from each simulator, reporting a failure only when there is consensus among the simulators.
In this method, the search process is conducted separately for each simulator, which can lead to numerous simulator-specific failures. Since discrepancies between simulators are only addressed after the searches terminate, this can lead to unnecessary consumption of the testing budget when such failures occur.

Following the approach of Biagiola et al.~\cite{matteoMaxibon2024}, we also base our methodology on multi-simulator, search-based testing, leveraging simulators capable of accommodating analogous configurations in terms of test scenarios and ADAS. However, our approach differs by performing a joint test evaluation \textit{during the search}, ensuring that scenarios leading to simulator-dependent failures (i.e., simulator disagreements) are filtered out before progressing further.

\section{Conclusion and Future Work}
\label{sec:conclusion}
In this paper, we presented the approach \msim, to mitigate simulator-specific failures when testing ADAS using an ensemble of simulators. 
Our approach leverages search-based testing and executes test inputs in multiple simulators during the search to identify generalizable failures.

In our empirical study, we evaluated our approach on testing two DNN-enabled ADAS trained for lane-keeping. We compared our approach in terms of effectiveness and efficiency in identifying simulator-agnostic/valid failures, to single-simulator based testing as well as state-of-the-art testing approach which employs multiple simulators.
The results of the study show that combining evaluation results from multiple simulators during testing outperforms single-simulator testing, identifying nearly 100\% valid failures. Compared to existing multi-simulator approaches, \msim identifies, on average, more valid failures. In terms of efficiency, \msim performs similarly to other approaches. Additionally, the study demonstrates that machine learning can be used within \msim to predict disagreements between test outcomes across different simulators, increasing efficiency and reducing the variation in the number of valid failures.

Our future work is to extend our study to other use cases including more complex systems under tests such as full-stack ADAS and to investigate the root causes behind the disagreements of evaluation outcomes. \changed{Another future direction could consist in the integration of the proposed multi-simulator concept within other existing test generators, which can act as a complementary layer that analyzes simulator-induced flakiness and verifies the transferability of detected faults across environments.}

\section{Declarations}  

\subsection{Funding}
This research was funded by the Bavarian Ministry of Economic Affairs, Regional Development and Energy. 
\blue{Matteo Biagiola is partially supported by Fondo Istituzionale per la Ricerca granted by Università della Svizzera italiana (USI).}

\subsection{Ethical Approval}  
Not applicable.

\subsection{Informed Consent}  
Not applicable.

\subsection{Author Contributions}  

\textbf{Lev Sorokin}: conceptualization, methodology, implementation, evaluation, writing, review, editing. \textbf{Matteo Biagiola}: conceptualization, methodology, review, editing. \textbf{Andrea Stocco}: conceptualization, methodology, review, editing.

\subsection{Data Availability Statement}  

All our results, the source code, and the simulator are accessible and can be reproduced~\cite{replication-package}.

\subsection{Conflict of Interest}  
The authors declare no conflict of interest.

\subsection{Clinical Trial Registration}  
Clinical trial number: Not applicable.

\bibliographystyle{spmpsci}
\bibliography{papers}

\end{document}